\documentclass[12pt,notitlepage]{revtex4-1}
\usepackage{graphicx}
\usepackage{subcaption}
\usepackage{amsfonts, amssymb, amsmath}
\usepackage{bbm}
\usepackage[utf8]{inputenc}
\usepackage[OT4]{fontenc}
\usepackage{url}
\usepackage{tensor}
\newcommand{\Hil}{\mathcal{H}}
\newcommand{\CubicalSymmetries}{O_{\rm cube}}

\newcommand{\CE}{{\hat{C}_E}}
\newcommand{\CL}{{\hat{C}_L}}

\newcommand{\ket}[1]{| #1 >}
\newcommand{\bra}[1]{< #1 | }

\newcommand{\orbit}[1]{\mathcal{O}_{#1}}
\newcommand{\stabilizer}[1]{\mathcal{S}_{#1}}
\newcommand{\AngleOperator}[1]{\hat{\theta}_{#1}}
\newcommand{\Angle}[1]{\theta_{#1}}
\newcommand{\spacevector}[3]{\begin{pmatrix}
#1\\#2\\#3
\end{pmatrix}}
\newcommand{\braket}[2]{\left< #1 | #2 \right>}

\newcommand{\iu}{{\it i}}

\newcommand{\sgn}{{\rm sgn}}

\newcommand{\link}{\ell}

\newcommand{\Inv}[1]{{\rm Inv}\left( #1 \right)}

\newcommand{\node}{n}

\newcommand{\valence}[1]{{\rm Valence}(#1)}

\newcommand{\id}{\mathbbm{1}}

\newcommand{\numberofloops}{\mathfrak{L}}
\newcommand{\loopconfiguration}{\mathfrak{l}}

\newcommand{\Tr}[1]{{\rm Tr}\left(#1 \right)}

\usepackage{accents}

\def\be{\begin{equation}}
\def\ee{\end{equation}}
\usepackage{color}

\renewenvironment{pmatrix}{\left(\begin{array}{ccc}}{\end{array}\right)}

\begin{document}
\title[Bouncing universe in LQG]{Bouncing universe in Loop Quantum Gravity: full theory calculation}
\author{Marcin Kisielowski}
\affiliation{ National Centre for Nuclear Research, Pasteura 7, 02-093 Warsaw, Poland}
\affiliation{ Centre de Physique Theorique, Aix-Marseille Univ., Luminy Case 907, 13288 Marseille, France}
%\address{National Centre for Nuclear Research, Pateura 7, 02-093 Warsaw, Poland}
\email{Marcin.Kisielowski@gmail.com}
\begin{abstract}
In Loop Quantum Gravity mathematically rigorous models of full quantum gravity were proposed. In this paper we will study a cosmological sector of one of the models describing quantum gravity with positive cosmological coupled to massless scalar field. In our previous research we introduced a method to reduce the model to homogeneous-isotropic sector at the quantum level. In this paper we propose a method to restrict to the spatially flat sector. After this restriction the number of degrees of freedom gets substantially reduced. This allows us to make numerical calculations. Remarkably, the resulting model shares some structural similarities with the Loop Quantum Cosmological models and therefore sheds some new light on the relation between Loop Quantum Gravity and Loop Quantum Cosmology. According to our model the evolution of the Universe is periodic. The quantum gravity effects resolve the Big Bang singularity leading to a Big Bounce and cause the Universe to contract after a classical expansion phase (Big Crunch). 
\end{abstract}
\maketitle
\section{Introduction}
Loop Quantum Gravity \cite{StatusReport,ThiemannBook,RovelliBook,Zakopane,LQG25,MaReview,AshtekarRovelliReuterRev} is rapidly developing research field. In particular, mathematically sound quantum gravity models were proposed  \cite{AQGIV,GravityQuantized} and the challenge is now to model physical phenomena. Remarkable results were obtained for symmetry reduced models, such as quantum resolution of the Big Bang singularity. In the early times most efforts were focused on quantizing models symmetry reduced at the classical level. This lead to a research field called Loop Quantum Cosmology \cite{LQCI,LQCII,LQCIII,LQCIV}. Recently much progress has been made to study the symmetry reduction at the quantum level  \cite{QRLQGI,QRLQGII,QRLQGIII,QRLQGIV,QRLQGV,QRLQGVI,QRLQGVII, CosmoGFTI,CosmoGFTII, DLI,DLII, HanEffectiveDynamics, HomogeneousIsotropic}. This paper continues our efforts in the second direction but, remarkably, our resulting model turns out to be similar to the LQC models.

In \cite{HomogeneousIsotropic} we proposed a reduction of degrees of freedom to homogeneous-isotropic geometries at the quantum level. We restricted the Hilbert space to a cubical lattice with loops -- in addition to the sides of the cubical lattice, we included also links starting and ending at the same node, called loops. Such loops are created and annihilated by the gravitational part of the scalar constraint. We constructed homogeneous-isotropic states by averaging over discrete translations and discrete rotations that are symmetries of the cubical lattice. The quantum Hamiltonian proposed in \cite{LewandowskiSahlmann,HamiltonianOperator,NewScalarConstraint,TimeEvolution} commutes with the symmetries and therefore descends to the symmetry reduced space. We introduced a cut-off in the number of loops in order to make numerical calculations. We observed that after the symmetry reduction the rank of the truncated matrix gets approximately $24$ times smaller. On the one hand, this opened a possibility of studying the model further using the same computing resources. On the other hand the reduction of the degrees of freedom was still very small compared to the drastic reduction at the classical level. In this paper we argue that at the quantum level a sector has to be chosen: a sector of positive, negative or zero intrinsic curvature. In this paper we will extract the zero intrinsic curvature sector. This restriction will be done approximately using the Livine-Speziale coherent states. It leads to substantial reduction of the degrees of freedom at the quantum level and allows numerical study of the model.

In this paper we will consider spin-networks in which all links are labelled with spin $\frac{1}{2}$. The quantum Hamiltonian from \cite{LewandowskiSahlmann,HamiltonianOperator,NewScalarConstraint,TimeEvolution} restricted to the spaces of monochromatic spin $\frac{1}{2}$ intertwiners has a property which makes it very similar to the quantum Hamiltonian in Loop Quantum Cosmology. In \cite{VolumeMonochromatic} we noticed that the Rovelli-Smolin volume operator \cite{RovelliSmolinVolume} restricted to the spaces of monochromatic spin $\frac{1}{2}$ intertwiners is proportional to identity:
\begin{equation}\label{eq:volume_monochromatic}
\hat{V}=\frac{\kappa_0}{8} \left(\frac{8\pi G \hbar \gamma}{c^3}\right)^{\frac{3}{2}} \sqrt{\frac{\sqrt{3}}{3!} (N-2)N(N+2)}\cdot \id.
\end{equation}
The proportionality factor depends only on the valence of the intertwiner space $N$. There are three terms in the gravitational part of the scalar constraint \cite{LewandowskiSahlmann,HamiltonianOperator,NewScalarConstraint,TimeEvolution} : one adding a loop, one removing a loop and one which does not change the graph. Adding a loop increases the valence of the intertwiner by $2$ and removing a loop decreases the valence by the same number. As a result, the gravitational part of the scalar constraint, in both Loop Quantum Cosmology and in Loop Quantum Gravity models proposed in \cite{LewandowskiSahlmann,HamiltonianOperator,NewScalarConstraint,TimeEvolution}, acting on a volume eigenvector produces a state which is a linear combination of three volume eigenstates: one with a higher volume, one with a lower volume and one with the same volume as the original state. This analogy will allows us to build a quantum Hamiltonian in our model which resembles the LQC Hamiltonian. The crucial difference between the approaches will be that our Hilbert space will be spanned by volume eigenvectors which do not correspond to equally spaced eigenvalues.
\section{Homogeneous-isotropic coherent states}\label{sc:coherent_states}
\begin{figure}[!tbp]
\centering
\centering
  \subfloat[A neighbourhood of the node $\node$ of a lattice with loops corresponding to $\loopconfiguration_\node$. This representation was used in \cite{HomogeneousIsotropic}]{\includegraphics[scale=0.75]{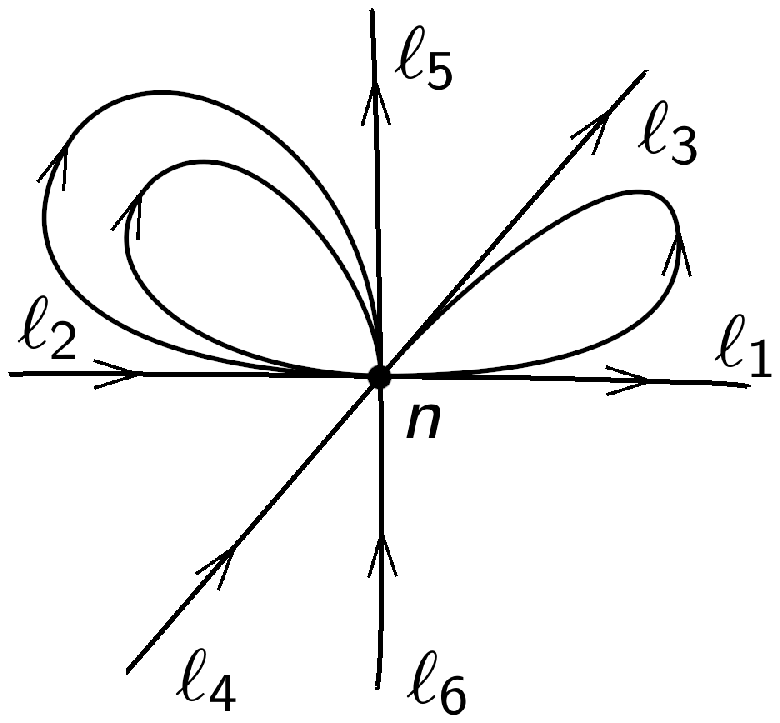}\label{fig:loop_configuration}}
  \hfill
  \subfloat[A neighbourhood of the node $\node$ of an equivalent lattice with loops corresponding to $\loopconfiguration$. This representation is used in this paper.]{\includegraphics[scale=0.75]{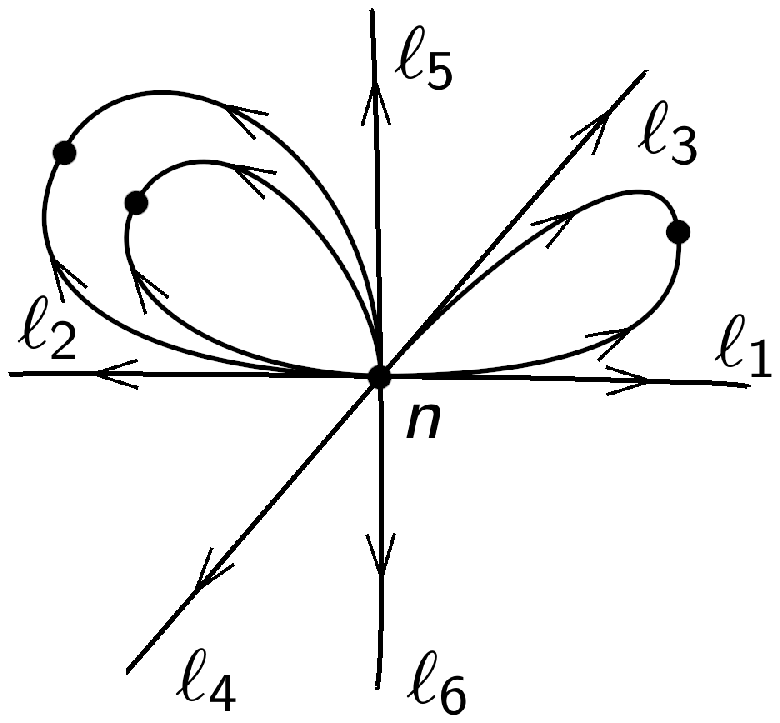}\label{fig:loop_configuration_divided}}
\caption{A loop configuration $\loopconfiguration_\node$ encodes the information about the loops tangent to the sides of the cubical lattice $\link_1, \ldots, \link_6$. On this figure we drew 2 equivalent neighbourhoods of a node  $\node$ corresponding to loop configuration $\loopconfiguration_\node$ such that $\loopconfiguration_\node(1,3)=1, \loopconfiguration_\node(2,5)=2, \loopconfiguration_\node(I,J)=0$ if $(I,J)\not\in\{(1,3),(2,5)\}$. In this paper we prefer to use the representation from figure (b). We denote by $\numberofloops(\loopconfiguration_\node)$ the total number of loops at the node $\node$. In this example $\numberofloops(\loopconfiguration_\node)=3$.}\label{fig:loop_configuration_ab}
\end{figure}

In \cite{HomogeneousIsotropic} we constructed Loop Quantum Gravity states that are homogeneous-isotropic. They form a subspace invariant under the action of the Hamiltonian. This subspace was build by considering a cubical lattice $\Gamma$ and adding loops (links starting and ending at a node) tangential to two different sides of the lattice.  Let us denote by $\link_1,\ldots, \link_6$ the links of the graph $\Gamma$ meeting at the node $n$ -- the six sides of the cubical lattice at the node $\node$ (see figure \ref{fig:loop_configuration_ab}). Let us consider a map:
\begin{equation}
\loopconfiguration_\node :\{(I,J): I<J, \epsilon(\dot{\link}_I, \dot{\link}_J)\neq 0\} \to \mathbb{N},
\end{equation}
where 
\begin{equation}\label{eq:epsilon}
\epsilon(\dot{\link}_I, \dot{\link}_J)=
\begin{cases}
0,&\textrm{ if the vectors }\dot{\link}_I, \dot{\link}_J\textrm{ are colinear},\\
1 & \textrm{otherwise}.
\end{cases}
\end{equation}
In the formula above we denoted by $\dot{\link}$ a vector tangent to link $\link$ at the node $\node$. We will call $\loopconfiguration_\node$ a loop configuration at the node $\node$. We will denote by $\numberofloops(\loopconfiguration_\node)$ the total number of loops at the node $\node$:
\begin{equation}
\numberofloops(\loopconfiguration_\node)=\sum_{I<J} \loopconfiguration_\node(I,J).
\end{equation}
The number $\loopconfiguration(I,J)$ is the number of loops tangential to both links $\link_I$ and $\link_J$ simultaneously. An example is given in figure \ref{fig:loop_configuration_ab}. Let us underline a subtle difference between the representations of the $\loopconfiguration_\node$ from \cite{HomogeneousIsotropic} (illustrated on figure \ref{fig:loop_configuration}) and the representation of $\loopconfiguration_\node$ in this paper (illustrated on figure \ref{fig:loop_configuration_divided}). In this paper we use an equivalent spin network (in the sense of cylindrical equivalence -- see for example \cite{StatusReport}) such that each loop is divided into two links by adding a bi-valent node and all links at the original nodes (at least $6$-valent) are outgoing. In \cite{HomogeneousIsotropic} we chose a spin $j$ labelling the sides of the cubical lattice and a spin $l$ labelling the loops. In this paper we limit to the case $j=l=\frac{1}{2}$ and we will denote by $\Hil_{\frac{1}{2}}$ the spin $\frac{1}{2}$ representation space of the SU(2) group. Let us notice that in the neighbourhood of the node $\node$ there are $\numberofloops(\loopconfiguration_\node)$ bivalent nodes and one node of valence $2\numberofloops(\loopconfiguration_\node)+6$ (see figure \ref{fig:loop_configuration_ab}). With the node $\node$ and the loop configuration $\loopconfiguration_\node$ we associate a space of intertwiners:
\begin{equation}
\Hil_{\loopconfiguration_\node}= \Inv{\underbrace{\Hil_{\frac{1}{2}}\otimes\ldots\otimes \Hil_{\frac{1}{2}}}_{2\numberofloops(\loopconfiguration_n)+6}}\otimes \underbrace{\Inv{\Hil_{\frac{1}{2}}^*\otimes \Hil_{\frac{1}{2}}^*}\otimes \ldots \otimes \Inv{\Hil_{\frac{1}{2}}^*\otimes \Hil_{\frac{1}{2}}^*} }_{ \numberofloops(\loopconfiguration)}.
\end{equation}
The Hilbert spaces $\Inv{\Hil_{\frac{1}{2}}^*\otimes \Hil_{\frac{1}{2}}^*}$ correspond to bivalent nodes and the Hilbert space 
\be 
\Inv{\underbrace{\Hil_{\frac{1}{2}}\otimes\ldots\otimes \Hil_{\frac{1}{2}}}_{2\numberofloops(\loopconfiguration_n)+6}}
\ee
corresponds to the node of the original cubical lattice (with loops).

In this paper we will fix a node of the cubical lattice $\node$ and omit the subscript $\node$ in $\loopconfiguration_{\node}$. As we argued in \cite{HomogeneousIsotropic} the diagonalization problem of the quantum Hamiltonian in the version proposed in \cite{HomogeneousIsotropic} can be considered at each node of the graph separately.  The eigenvalues of the full operator are tensor products of eigenvalues at each node. The quantum Hamiltonian at a fixed node $n$ can be considered as an operator in a Hilbert space $\Hil^{\rm loop}_{\Gamma,j,n}$ which is a direct product of the spaces of intertwiners corresponding to different loop configurations:
\begin{equation}\label{eq:Hilbert_space}
\Hil^{\rm loop}_{\Gamma,j,n}=\bigoplus_{\loopconfiguration} \Hil_{\loopconfiguration}.
\end{equation}
A convenient basis of our space is of the form
\begin{equation}
\ket{\loopconfiguration,\iota},
\end{equation}
where $\loopconfiguration$ runs through set of loop configurations and $\iota$ runs through a basis of the intertwiners $\Hil_{\loopconfiguration}$. In particular:
\begin{equation}
\braket{\loopconfiguration',\iota'}{\loopconfiguration,\iota}=\delta_{\loopconfiguration',\loopconfiguration} \braket{\iota'}{\iota}.
\end{equation}

\subsection{Coherent intertwiners}\label{sc:coherent_intertwiners}
Our coherent states are built from Livine-Speziale coherent states. In this paper we will take $j=1/2$. There are 2 types of nodes in the graph: the nodes that are also nodes of the lattice with loops (they are at least 6-valent) and the nodes that are obtained by splitting loops into two links (they are at 2-valent). For each $\loopconfiguration$ we will construct a tensor $\iota_\loopconfiguration$ which is a tensor product of some specific intertwiners at the 2-valent nodes and a specific intertwiner at the at least $6$-valent node $\node_0$:
\be\label{eq:coherent_intertwiners}
\iota_\loopconfiguration = \left(\bigotimes_{\node: \valence{\node}=2} \iota_\node^\dagger \right) \otimes \iota_{\node_0}
\ee
We will call $\iota_\loopconfiguration$ a coherent intertwiner. In the next subsections we will describe the intertwiners $\iota_\node$ and $\iota_{\node_0}$.
\subsubsection{At least 6-valent node}
Let us consider a small neighbourhood of at least 6-valent node $\node_0$ of the graph. This neighbourhood is isomorphic with a small neighbourhood of the graph $\widetilde{\Gamma}$ in $\mathbb{R}^3$. Therefore the possible directions of vectors tangent to the links of the graph at the node $\node_0$ are $\partial_x,-\partial_x,\partial_y,-\partial_y,\partial_z,-\partial_z$. To each link $\link$ at the node $\node_0$ there corresponds a Perelomov coherent state $\ket{1/2,\dot{\link}}$, where $\dot{\link}$ is the vector tangent to $\link$ at the node $\node$. Therefore the tensor assigned to the node $\node_0$ is defined by Livine-Speziale coherent intertwiner:
\be
\ket{\loopconfiguration, \vec{n}_1 \ldots \vec{n}_6}:= \int_{\rm SU(2)} du\ \rho_{1/2}(u)^{\otimes \numberofloops(\loopconfiguration)} \bigotimes_{I=1}^6 \ket{1/2,\vec{n}_I}^{\otimes N_I},
\ee
where $N_I=\sum_{J=1}^{I-1} \loopconfiguration(J,I)+\sum_{J=I+1}^6 \loopconfiguration(I,J)+1$ is the number of links starting at the node $\node_0$ whose tangent direction is $\vec{n}_I$,
\be
\vec{n}_1=\spacevector{1}{0}{0},\ \vec{n}_2=\spacevector{-1}{0}{0},\ \vec{n}_3=\spacevector{0}{1}{0},\ \vec{n}_4=\spacevector{0}{-1}{0},\ \vec{n}_5=\spacevector{0}{0}{1},\ \vec{n}_6=\spacevector{0}{0}{-1}.
\ee
In our notation,
\be 
\ket{1/2,\vec{n}_I}^{\otimes N_I}:=\underbrace{\ket{1/2,\vec{n}_I}\otimes \ldots \otimes \ket{1/2,\vec{n}_I}}_{N_I}.
\ee
It is straightforward to notice that $\ket{1/2,\vec{n}}^{\otimes N}$ is the same state as the Perelomov coherent state with spin $N/2$ and direction vector $\vec{n}$ (under the standard isomorphism mapping spinors into the corresponding states in the magnetic number basis):
\be\label{eq:spinors_iso_magnetic}
\ket{1/2,\vec{n}}^{\otimes N}=\ket{N/2,\vec{n}}.
\ee
Under this isomorphism we have:
\be\label{eq:loop_state_isomorphism}
\ket{\loopconfiguration, \vec{n}_1 \ldots \vec{n}_6} = \ket{j_1,\ldots, j_6; \vec{n}_1 \ldots \vec{n}_6},
\ee
where $j_i=N_i/2$ and $\ket{j_1,\ldots, j_6/2; \vec{n}_1 \ldots \vec{n}_6}$ is the Livine-Speziale coherent intertwiner:
\be
\ket{j_1,\ldots, j_6; \vec{n}_1 \ldots \vec{n}_6}=\int_{\rm SU(2)} du\ \bigotimes_{I=1}^6 \rho_{j_I}(u)  \ket{j_I,\vec{n}_I}.
\ee

We will assign to the node $\node_0$ the normalized Livine-Speziale intertwiner:
\be
\iota_{\node_0}=\frac{1}{\sqrt{\braket{\loopconfiguration, \vec{n}_1 \ldots \vec{n}_6 }{\loopconfiguration, \vec{n}_1 \ldots \vec{n}_6}}}\ket{\loopconfiguration, \vec{n}_1 \ldots \vec{n}_6}.
\ee
\subsubsection{2-valent nodes}
The 2-valent nodes are added to the graph to split the loops of our lattice with loops. As a result, the node is an endpoint of 2 links which are tangent to 2 different sides of the lattice $\link_I$ and $\link_J$. We can associate with such node an intertwiner $\iota_\node^\dagger$, where $\iota_\node$ is a normalized Livine-Speziale coherent intertwiner. Strictly speaking, we consider a Livine-Speziale coherent intertwiner:
\be
\ket{1/2,1/2; \vec{n}_I,\vec{n}_J}:= \int_{\rm SU(2)} du\ \rho_{1/2}(u)\otimes \rho_{1/2}(u)\  \ket{1/2,\vec{n}_I}\otimes \ket{1/2,\vec{n}_J}.
\ee
In the formula above the vectors $\vec{n}_I$ and $\vec{n}_J$ are 2 vectors parallel to the sides of the lattice $\link_I$ and $\link_J$. We fix the ambiguity in the definition of the state by requiring that $I<J$. As previously, we will associate with the node an intertwiner $\iota_\node$
\be
\iota_{\node}=\frac{1}{\sqrt{\braket{1/2,1/2; \vec{n}_I,\vec{n}_J}{1/2,1/2; \vec{n}_I,\vec{n}_J}}}\ket{1/2,1/2; \vec{n}_I,\vec{n}_J}.
\ee

Let us notice that any element of the space $\Inv{\Hil_{1/2}\otimes \Hil_{1/2}}$ is proportional to $\frac{1}{\sqrt{2}}\epsilon^{AB}$. Since $\iota_{\node}$ is normalized, it can only differ from $\frac{1}{\sqrt{2}}\epsilon^{AB}$ by phase. Let us denote by \be
\epsilon(\vec{n}_I,\vec{n}_J)=\epsilon_{AB} \vec{n}_I^A \vec{n}_J^B, 
\ee 
where $A,B\in\{-\frac{1}{2},\frac{1}{2}\}$ are spinor indices. Using this notation
\be\label{eq:two_valent_intertwiner}
\iota_n^{AB} = \frac{1}{\sqrt{2}} \frac{\epsilon(\vec{n}_I,\vec{n}_J)}{|\epsilon(\vec{n}_I,\vec{n}_J)|} \epsilon^{AB} .
\ee
For completeness, let us calculate $|\epsilon(\vec{n}_I,\vec{n}_J)|$:
\begin{multline}
 |\epsilon(\vec{n}_I,\vec{n}_J)|^2=\overline{\vec{n}_I^A \epsilon_{AB}  \vec{n}_J^B}\ \vec{n}_I^{A'} \epsilon_{A'B'} \vec{n}_J^{B'}=-\overline{\vec{n}_I^A \epsilon_{AB}  \vec{n}_J^B}\ \vec{n}_J^{A'} \epsilon_{A'B'} \vec{n}_I^{B'}=\\
 =-\Tr{\ket{1/2,\vec{n}_I}\bra{1/2,\vec{n}_I}\ \epsilon\ \left(\ket{1/2,\vec{n}_J}\bra{1/2,\vec{n}_J} \right)^T\ \epsilon}=\\=-\frac{1}{4}\Tr{(\id+\vec{n}_I\cdot \vec{\sigma})\epsilon (\id+\vec{n}_J\cdot \vec{\sigma}^T)\epsilon}=-\frac{1}{4}\Tr{(\id+\vec{n}_I\cdot \vec{\sigma})(-\id+\vec{n}_J\cdot \vec{\sigma})}=-\frac{1}{4}\left(-2+2\, \vec{n}_I \cdot \vec{n}_J \right)=\frac{1}{2}.
\end{multline}
Inserting this result into \eqref{eq:two_valent_intertwiner} we obtain:
\be
\iota_n^{AB} = \epsilon(\vec{n}_I,\vec{n}_J) \epsilon^{AB}.
\ee

\subsection{Invariance of the coherent intertwiners}
The action of the group of orientation preserving isometries of a cube $\CubicalSymmetries$ induces an action of the permutation group on the links of the graph. This permutation induces an action of the group on loop configurations $\loopconfiguration$ and on intertwiners. The states transform in the following way:
\be\label{eq:coherent_intertwiners_transformation}
U_g  \ket{\loopconfiguration, \iota_\loopconfiguration}= \sgn(g)\,\ket{g\cdot\loopconfiguration, R(g) \iota_\loopconfiguration},
\ee
where the action of $g$ on $\loopconfiguration$ was described in \cite{HomogeneousIsotropic} and $R(g)$ acts by permuting the indices of the intertwiners:
\be\label{eq:permutation_of_indices}
R(g)\ket{j_1, \ldots j_6; \vec{n_1} \ldots \vec{n_6}}= \ket{j_{\sigma_g^{-1}(1)}, \ldots j_{\sigma_g^{-1}(6)};\vec{n}_{\sigma_g^{-1}(1)} \ldots \vec{n}_{{\sigma}_g^{-1}(6)}}.
\ee
 We will show now that
\be
U_g  \ket{\loopconfiguration, \iota_\loopconfiguration}=\ket{g\cdot\loopconfiguration, \iota_{g\cdot\loopconfiguration}}.
\ee
In the proof the fundamental role is played by the transformation property of the Perelomov coherent states under the action of an SU(2) group:
\be\label{eq:transformation_of_Perelomov_states}
\rho_j(u) \ket{j,\vec{n}}=e^{\iu\, \Phi(j,\vec{n},u)} \ket{j,u\cdot \vec{n}},
\ee
where $u\cdot \vec{n}$ is the unit vector obtained by rotating the vector $\vec{n}$ with an SO(3) group element corresponding to $u$, the phase $\Phi(j,\vec{n},u)$ is given in \cite{Perelomov1986} and its particular form will not be used explicitly here. 

Let us consider the transformation property of the intertwiner at the node $\node_0$. In this case we can write the transformation \eqref{eq:permutation_of_indices}  in the following form:
\be
R(g)\ket{\loopconfiguration; \vec{n_1} \ldots \vec{n_6}}=\ket{g\cdot \loopconfiguration; \vec{n}_{\sigma_g^{-1}(1)} \ldots \vec{n}_{{\sigma}_g^{-1}(6)}}=\ket{g\cdot \loopconfiguration; g\cdot \vec{n}_{{1}} \ldots g\cdot \vec{n}_{{6}}}.
\ee
From the property \eqref{eq:transformation_of_Perelomov_states} and from the invariance of the Haar measure it follows that
\be\label{eq:6_valent_node_phase}
\ket{g\cdot \loopconfiguration; g\cdot \vec{n}_{{1}} \ldots g\cdot \vec{n}_{{6}}}\sim  \ket{g\cdot \loopconfiguration; \vec{n}_{{1}} \ldots \vec{n}_{{6}}},
\ee
where $\sim$ denotes proportionality up to a phase.

Let us consider the transformation property of an intertwiner at a bivalent node $\node$ splitting a loop into two links.
In this case
\be
R(g) \ket{1/2,1/2;\vec{n}_I,\vec{n}_J}=\begin{cases} \ket{1/2,1/2;\vec{n}_J,\vec{n}_I},& \textrm{ if $g$ flips the loop}, \\
\ket{1/2,1/2;\vec{n}_I,\vec{n}_J},& \textrm{otherwise}.
\end{cases}
\ee
After the action of the group element $g$, the loop is between links $I'$ and $J'$, where:
\be
\sigma_g(I)=I',\quad \sigma_g(J)=J'.
\ee
There are 2 cases:
\begin{itemize}
\item $I'<J'$. In this case
\begin{multline}\label{eq:2_valent_node_phase_I}
R(g) \ket{1/2,1/2;\vec{n}_I,\vec{n}_J}= \ket{1/2,1/2;\vec{n}_I,\vec{n}_J}=\\=\ket{1/2,1/2;\vec{n}_{\sigma_{g^{-1}}(I')},\vec{n}_{\sigma_{g^{-1}}(J')}}\sim \ket{1/2,1/2;\vec{n}_{I'},\vec{n}_{J'}}.
\end{multline}
\item $I'>J'$. In this case
\begin{multline}\label{eq:2_valent_node_phase_II}
R(g) \ket{1/2,1/2;\vec{n}_I,\vec{n}_J}= \ket{1/2,1/2;\vec{n}_J,\vec{n}_I}=\\=\ket{1/2,1/2;\vec{n}_{\sigma_{g^{-1}}(J')},\vec{n}_{\sigma_{g^{-1}}(I')}}\sim \ket{1/2,1/2;\vec{n}_{J'},\vec{n}_{I'}}.
\end{multline}
\end{itemize}

The phases in \eqref{eq:6_valent_node_phase},\eqref{eq:2_valent_node_phase_I},\eqref{eq:2_valent_node_phase_II} can be calculated using formula \eqref{eq:transformation_of_Perelomov_states}. Since 2-valent intertwiners come with dagger into the definition of the coherent state \eqref{eq:coherent_intertwiners}, there is substantial cancellation of phases when $R(g)$ is applied. As a result, to the overall phase contribute only the Perelomov coherent states that correspond to the sides of the lattice  and
\be
R(g)\iota_{\loopconfiguration}=\sgn(g) \iota_{g\cdot\loopconfiguration}.
\ee
Combining this result with the transformation property of our coherent states \eqref{eq:coherent_intertwiners} gives the covariance property of our states:
\be\label{eq:transformation_property_full_state}
U_g  \ket{\loopconfiguration, \iota_\loopconfiguration}= \ket{g\cdot\loopconfiguration, \iota_{g\cdot\loopconfiguration}},
\ee
It will be convenient to introduce a notation
\be
\ket{\loopconfiguration}:=\ket{\loopconfiguration, \iota_\loopconfiguration}.
\ee
Let us notice that from the transformation property \eqref{eq:transformation_property_full_state} it follows that a state $\ket{\loopconfiguration}$ is invariant under the the stabilizer group of $\loopconfiguration$.

\subsection{Homogeneous-isotropic states}
Our homogeneous-isotropic states are obtained by averaging states $\ket{\loopconfiguration}$ over the group of orientation preserving symmetries of a cube $\CubicalSymmetries$. Let us denote by $\orbit{\loopconfiguration}$ the orbit of the action of the group $\CubicalSymmetries$ on a loop configuration $\loopconfiguration$. The homogeneous-isotropic state corresponding to $\ket{\loopconfiguration}$ is:
\be
\ket{[\loopconfiguration]}=\frac{1}{\sqrt{\# \orbit{\loopconfiguration}}}\  \sum_{\loopconfiguration'\in \orbit{\loopconfiguration}}\ \ket{\loopconfiguration'}.
\ee
Thanks to the numerical factor in front, the states are normalized. Let us denote by $\stabilizer{\loopconfiguration}$ the stabilizer group of $\CubicalSymmetries$ with respect to $\loopconfiguration$. The formula above can be written using the group action explicitly:
\be\label{eq:hom_iso_states_group}
\ket{[\loopconfiguration]}=\frac{1}{\sqrt{\# \orbit{\loopconfiguration}}\# \stabilizer{\loopconfiguration}}\  \sum_{g\in \CubicalSymmetries}\ U_g \ket{\loopconfiguration}=\frac{1}{\sqrt{24\cdot \# \stabilizer{\loopconfiguration}}}\  \sum_{g\in \CubicalSymmetries}\ U_g \ket{\loopconfiguration}.
\ee

\subsection{The basis states}\label{sc:basis_states}
We will further restrict the space to states which are defined by the number of loops only
\be
\ket{k}:=\ket{[\loopconfiguration_k]},
\ee
where $\loopconfiguration_k$ is loop configuration defined in the following way. Let us split $k$ into a quotient $n$ and a remainder $r$ with respect to division by $3$:  $k=3\,n+r$, where $n,r\in \mathbb{N}$ and $r<3$. We consider three cases:
\begin{itemize}
\item $r=0$. In this case:
\be
\loopconfiguration_k(1,3)=\loopconfiguration_k(2,6)=\loopconfiguration_k(4,5)=n.
\ee
\item $r=1$. In this case:
\be
\loopconfiguration_k(1,3)=n+1,\quad \loopconfiguration_k(2,6)=\loopconfiguration_k(4,5)=n.
\ee
\item $r=2$. In this case:
\be
\loopconfiguration_k(1,3)=\loopconfiguration_k(2,6)=n+1,\quad \loopconfiguration_k(4,5)=n.
\ee
\end{itemize}
All the remaining $\loopconfiguration(I,J)$ are $0$. 

The space of such states will be denoted by $\Hil_{\rm flat}$.
\section{Internal flatness of the quantum geometry}\label{sc:internal_flatness}
In order to investigate the internal geometry of the states proposed in the previous section, we will study the expectation values of the area and angle operators. We will consider a small cube whose center is at the node $\node_0$ and such that:
\begin{itemize}
\item the outward pointing normals to the faces of the cube are $\vec{n}_1,\ldots, \vec{n}_6$,
\item each link intersects (transversally) only one face of the cube -- the face with the outward normal pointing in the same direction as the vector tangent to the link at the node $\node_0$.
\end{itemize}
 We will denote the faces of the cube by $S_I, I=1,\ldots,6$.

We will argue that the homogeneous-isotropic states proposed in \cite{HomogeneousIsotropic} have the property that the expectation values of the area operator is the same for any face of the cube and the dihedral angle operator is the same for any pair of non-parallel faces of the cube. For the states proposed in this paper, this dihedral angle will be approximately $\frac{\pi}{2}$. This will allows us to interpret the space of states considered in this paper as a subspace of homogeneous-isotropic states corresponding to internally flat geometries.
\subsection{Areas}
We refer our reader to \cite{StatusReport} for a detailed presentation of the area operator in Loop Quantum Gravity. In this paper we consider the particular case described in the introductory part of this section. Let us denote by $\link_I, I=\{1,\ldots,6\}$ the segments of the lattice (without loops) that intersect at $\node_0$. The area operator corresponding to the surface $S_I$ is:
\begin{equation}
\hat{A}_I=\frac{8\pi G}{c^3} \sqrt{ \hat{P}^i(S_I)  \hat{P}^j(S_I) \delta_{ij} }=4\pi \gamma l_P^2\sqrt{-\hat{J}^i_I \hat{J}^j_I\, \delta_{ij}},
\end{equation}
where 
\begin{equation}
\hat{J}^i_I=\sum_{r:\dot{\link}_r \parallel \dot{\link}_I} J^i_r.
\end{equation}
The expectation value of the area operator corresponding to a face $S_I$ is
\begin{equation}
A_I(\loopconfiguration):=\bra{[\loopconfiguration]} \hat{A}_I \ket{[\loopconfiguration]}=4\pi \gamma l_P^2 \bra{[\loopconfiguration]} \sqrt{-\hat{J}^i_I \hat{J}^j_I\, \delta_{ij}} \ket{[\loopconfiguration]}.
\end{equation}
Due to the invariance of the states this expectation value does not depend on $I$:
\begin{equation}
\forall_{g\in \CubicalSymmetries}\ A_{\sigma_g^{-1}(I)}(\loopconfiguration)= \bra{[\loopconfiguration]} \hat{A}_{\sigma_g^{-1}(I)} \ket{[\loopconfiguration]}=\bra{[\loopconfiguration]} U_g^\dagger \hat{A}_{I} U_g \ket{[\loopconfiguration]} = \bra{[\loopconfiguration]} \hat{A}_{I} \ket{[\loopconfiguration]}=A_{I}(\loopconfiguration).
\end{equation}
We will therefore introduce a notation:
\begin{equation}
A(\loopconfiguration):=A_{I}(\loopconfiguration).
\end{equation}
Let us notice that the expectation value of the area operator in the state $\ket{[\loopconfiguration]}$ can be expressed in terms of an expectation value of the area operator in the state $\ket{\loopconfiguration}$:
\begin{multline}
A(\loopconfiguration)=\bra{[\loopconfiguration]} \hat{A}_I \ket{[\loopconfiguration]}=\frac{1}{24\cdot \# \stabilizer{\loopconfiguration}} \sum_{g,g'\in \CubicalSymmetries} \bra{\loopconfiguration}  U_g^\dagger \hat{A}_I U_{g'}\ket{\loopconfiguration}=\frac{1}{24}\sum_{g\in \CubicalSymmetries} \bra{\loopconfiguration}  U_g^\dagger \hat{A}_I U_{g}\ket{\loopconfiguration}=\\=\frac{1}{6}\left(\sum_{I=1}^6 \sqrt{j_I(j_I+1)}\right).
\end{multline}
The last equality comes from the fact that $\hat{A}_I$ commutes with any transformation that does not change the spin label $j_I$, in particular with any element of $\CubicalSymmetries$ that leaves $\vec{n}_I$ invariant ($\#\stabilizer{\vec{n}_I}=4$).
\subsection{Angles}
We will use the (dihedral) angle operator defined in \cite{MajorQuantizedDirection,MajorSeifert}. For each pair $(I,J)$ such that $\vec{n}_I \cdot \vec{n}_J=0$ we consider an operator corresponding to a dihedral angle between the surfaces $S_I$ and $S_J$:
\begin{equation}
\cos \AngleOperator{IJ}= \frac{\hat{P}^i(S_I)  \hat{P}^j(S_J) \delta_{ij}}{ \sqrt{\hat{P}^i(S_I)  \hat{P}^j(S_I) \delta_{ij}}\sqrt{\hat{P}^i(S_J)  \hat{P}^j(S_J) \delta_{ij}} }
\end{equation}
It can be expressed using the angular momentum operators:
\begin{equation}
\cos \AngleOperator{IJ}= \frac{-\hat{J}^i_I \hat{J}^j_J \delta_{ij}}{ \sqrt{-\hat{J}^i_I  \hat{J}^j_I \delta_{ij}}\sqrt{-\hat{J}^i_J  \hat{J}^j_J \delta_{ij}} }.
\end{equation}
Due to the invariance of the states under $\CubicalSymmetries$ transformations the expectation value
\begin{equation}
\bra{[\loopconfiguration]} \cos \AngleOperator{IJ} \ket{[\loopconfiguration]}
\end{equation}
does not depend on $I,J$. Indeed,
\begin{equation}
\forall_{g\in \CubicalSymmetries}\ \bra{[\loopconfiguration]} \cos \AngleOperator{\sigma_g^{-1}(I) \sigma_g^{-1}(J)} \ket{[\loopconfiguration]}=\bra{[\loopconfiguration]} U_g^\dagger \cos \AngleOperator{I J} U_g \ket{[\loopconfiguration]} = \bra{[\loopconfiguration]} \cos \AngleOperator{I J} \ket{[\loopconfiguration]}.
\end{equation}
This means that the dihedral angle is the same between any pair of non-parallel faces $S_I, S_J$ and is a clear sign of isotropy of the quantum geometry. The expectation value of the dihedral angle operator in the state $\ket{[\loopconfiguration]}$ can be expressed in terms of an expectation value of the dihedral operator in the state $\ket{\loopconfiguration}$: 
\begin{multline}
\bra{[\loopconfiguration]} cos \AngleOperator{I J} \ket{[\loopconfiguration]}=\frac{1}{24\cdot \# \stabilizer{\loopconfiguration}} \sum_{g,g'\in \CubicalSymmetries} \bra{\loopconfiguration}  U_g^\dagger  \cos \AngleOperator{I J}  U_{g'}\ket{\loopconfiguration}=\frac{1}{24}\sum_{g\in \CubicalSymmetries} \bra{\loopconfiguration}  U_g^\dagger  \cos \AngleOperator{I J}  U_{g}\ket{\loopconfiguration}=\\=\frac{1}{12}\left(\sum_{I,J} \bra{\loopconfiguration} \cos \AngleOperator{I J} \ket{\loopconfiguration} \right).
\end{multline}
Let us introduce a notation:
\begin{equation}
\cos \Angle{ }\ (k):=\bra{k}\cos \AngleOperator{I J}\ket{k}.
\end{equation}
We calculated the expectation values of the cosine of the dihedral angle operator $\bra{k}\cos \AngleOperator{I J}\ket{k}$ for different values of number of loops $k$. The values grow and seem to reach zero asymptotically. The fact that the asymptotic value is zero can be calculated using the (extended) stationary phase method. This calculation will be done in the next subsection. 
\begin{figure}
\centering
\includegraphics{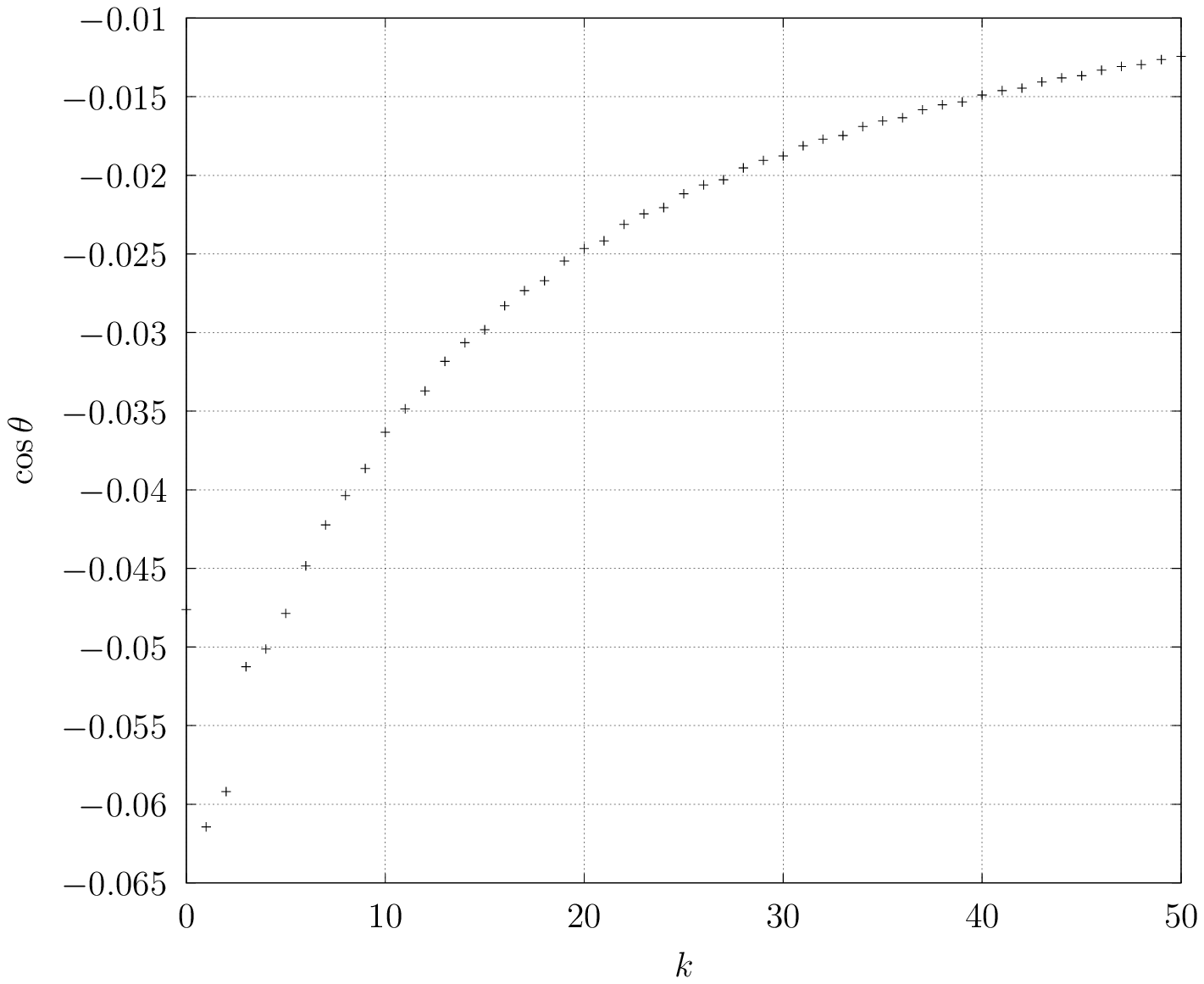}
\caption{Expectation values of the operator $\bra{k}\cos\AngleOperator{I J}\ket{k}$ as a function of $k$.}\label{fig:dihedral_angle}
\end{figure}
\subsection{Large Volume limit}\label{sc:flatness_large_volume}
The importance of the large volume limit comes from the fact that current Universe has a very large volume in the Planck units. This fact has been used for example in \cite{BRV} to study a semiclassical limit of a quantum cosmological model based on a spin-foam theory. In our approach, the volume operator is the Rovelli-Smolin volume operator, which for spin $1/2$ $N$-valent intertwiners is proportional to the identity. In \cite{VolumeMonochromatic} we have shown that:
\be
V_{\vec{j}}=\frac{\kappa_0}{8} \left(\frac{8\pi G \hbar \gamma}{c^3}\right)^{\frac{3}{2}} \sqrt{\frac{\sqrt{3}}{3!} (N-2)N(N+2)}\cdot \id,
\ee
where $\vec{j}=(\underbrace{\frac{1}{2},\ldots,\frac{1}{2}}_{N})$ encodes the spin labels of the $N$-valent intertwiner.

In this paper, the valence $N$ of an intertwiner at the node $\node_0$ is
\be
N=2k+6,
\ee
where $k$ is the number of loops at $\node_0$. Therefore, the large volume limit coincides with the limit of large number of loops. We will therefore consider the limit of large $n$, where the number of loops is $k=3n+r$. As we argued earlier, the intertwiners at the node $\node_0$ can be described by Livine-Speziale coherent states 
\be
\ket{j_1,\ldots,j_6;\vec{n}_1,\ldots,\vec{n}_6},
\ee
where $j_I=N_I/2$ and $N_I$ is the number of links starting at $\node_0$ whose tangent direction is $\vec{n}_I$ (see \eqref{eq:loop_state_isomorphism}). Due to the particular choice of loop configurations in the definition of the states we have the following values of spins:
\begin{itemize}
\item for $k=3n$ the spins are $j_1=\ldots=j_6=\frac{n+1}{2}$,
\item for $k=3n+1$ the spins are $j_1=j_3=\frac{n+2}{2}, j_2=j_4=j_5=j_6=\frac{n+1}{2}$,
\item for $k=3n+2$ the spins are $j_1=j_2=j_3=j_6=\frac{n+2}{2}, j_4=j_5=\frac{n+1}{2}$.
\end{itemize}
Therefore, the large volume limit translates into the limit of large spins $j_I$.

We will calculate the expectation value
\begin{equation}
\bra{\loopconfiguration_k}   \cos \AngleOperator{IJ} \ket{\loopconfiguration_k}=\bra{\loopconfiguration_k}   \frac{-\hat{J}^i_I \hat{J}^j_J \delta_{ij}}{ \sqrt{-\hat{J}^i_I  \hat{J}^j_I \delta_{ij}}\sqrt{-\hat{J}^i_J  \hat{J}^j_J \delta_{ij}} } \ket{\loopconfiguration_k}=  \frac{-\delta_{jl}\beta^{jl}_{IJ}}{\sqrt{j_I(j_I+1)}\, \sqrt{j_J(j_J+1)}\,N_k^2},
\end{equation}
where $N_k$ is the factor coming from normalization of the Livine-Speziale coherent intertwiners
\be\label{eq:Nk}
N_k = \sqrt{ \braket{\loopconfiguration_{k}, \vec{n}_1\ldots \vec{n}_6}{\loopconfiguration_{k}, \vec{n}_1\ldots \vec{n}_6}  }
\ee
and the remaining factor is
\begin{multline}
\beta^{jl}_{IJ}=\int_{\rm SU(2)} du \bra{j_I,\vec{n}_I}\hat{J}_I^j\, \rho_{j_I}(u) \ket{j_I,\vec{n}_I} \bra{j_J,\vec{n}_J}\hat{J}_J^l\, \rho_{j_J}(u) \ket{j_J,\vec{n}_J}\cdot \\ \cdot \prod_{K\not\in \{I,J\}}\bra{j_K,\vec{n}_K}\rho_{j_K}(u) \ket{j_K,\vec{n}_K}.
\end{multline}
Let us notice that 
\be\label{eq:beta_jl_IJ}
\beta^{jl}_{IJ} =\int_{\rm SU(2)} du\ B^{jl}_{IJ}(u) \prod_{K=1}^6\bra{j_K,\vec{n}_K}\rho_{j_K}(u) \ket{j_K,\vec{n}_K},
\ee
where
\be
B^{jl}_{IJ}(u)=j_I j_J \frac{\bra{1/2,\vec{n}_I}\tau^j \rho_{1/2}(u)\ket{1/2,\vec{n}_I}}{\bra{1/2,\vec{n}_I} \rho_{1/2}(u)\ket{1/2,\vec{n}_I}} \frac{\bra{1/2,\vec{n}_J}\tau^l \rho_{1/2}(u)\ket{1/2,\vec{n}_J}}{\bra{1/2,\vec{n}_J} \rho_{1/2}(u)\ket{1/2,\vec{n}_J}}.
\ee
The expression \eqref{eq:beta_jl_IJ} can be written in the following form:
\be
\beta^{jl}_{IJ} =\int_{\rm SU(2)} du\ B^{jl}_{IJ}(u) e^{S_{\vec{j}}(u)},
\ee
where
\be
S_{\vec{j}}(u)=\sum_{K=1}^6 2 j_K \ln \left(\bra{1/2,\vec{n}_K}\rho_{1/2}(u) \ket{1/2,\vec{n}_K}\right).
\ee
Let us consider the case $k=3n$. We scale $j_I=\frac{n+1}{2}$ by an overall constant $\lambda$: $j_I\mapsto \lambda j_I$. Under this scaling, the expression \eqref{eq:beta_jl_IJ} transforms as
\be
(\beta_\lambda)^{jl}_{IJ} =\lambda^2 \int_{\rm SU(2)} du\ B^{jl}_{IJ}(u)\, e^{\lambda S_{\vec{j}}(u)}.
\ee
In order to evaluate the integral, we will use the extended saddle point analysis (see for example \cite{EuclideanEPRLAsymptotics}). We notice, that the integral
\be
\int_{\rm SU(2)} du\ e^{S_{\vec{j}}(u)}
\ee
is the norm of the Livine-Speziale intertwiner $\ket{j_1,\ldots,j_6;\vec{n}_1,\ldots,\vec{n}_6}$, where $j_1=\ldots=j_6=\frac{n+1}{2}$. In this case the saddle point analysis has been performed in \cite{LSiI} and we can use the result from this paper to conclude that the only critical point of $S_{\vec{j}}$ is $u=\id$. In the saddle point approximation \cite{LSiI}
\be
{\beta_\lambda}^{jl}_{IJ} \approx\lambda^2 B^{jl}_{IJ}(\id)\frac{1}{\sqrt{\pi \lambda^{3} \det H}},
\ee
where $H$ is the Hessian of $S_{\vec{j}}$. Its explicit form will not be important here, because it gets cancelled with the Hessian from $\frac{1}{{N_\lambda}_k^2}$:
\be
\frac{{\beta_\lambda}^{jl}_{IJ}}{{N_\lambda}_k^2}\approx \lambda^2 B^{jl}_{IJ}(\id).
\ee
It is straightforward to calculate $B^{jl}_{IJ}(\id)$:
\be 
B^{jl}_{IJ}(\id)=-j_I j_J n^j_I n^l_J.
\ee
Since $\vec{n}_I\cdot \vec{n}_J=0$:
\begin{equation}
\delta_{ij}B^{jl}_{IJ}(\id)=0.
\end{equation}
We conclude that
\be
\bra{\loopconfiguration_k}   \cos \AngleOperator{IJ} \ket{\loopconfiguration_k} \to 0
\ee
in the large volume limit (the limit of large number of loops). The cases $k=3n+1$ and $k=3n+2$ are analogous. As a result the faces $S_I$ and $S_J$ are asymptotically orthogonal and the deficit angle (in the sense of Regge) is $0$. Therefore the geometries are flat.

\section{The Euclidean part of the Hamiltonian operator}\label{sc:euclidean_part}
We will start this section with recalling some properties of the quantum Hamiltonian operator. We will use the version of the operator from \cite{TimeEvolution} that corresponds to the model of quantum gravity coupled to massless scalar field. In particular, we will see that thanks to the results from \cite{VolumeMonochromatic} about the Rovelli-Smolin volume operator in the space of spin $\frac{1}{2}$ monochromatic intertwiners and due to internal flatness of the quantum geometries (which we discuss in section \ref{sc:internal_flatness}) the only non-trivial part is the Euclidean part of gravitational scalar constraint operator. We will study it in detail in this section.
\subsection{Quantum Hamiltonian operator}
For the class of operators considered in \cite{LewandowskiSahlmann,HamiltonianOperator,NewScalarConstraint,TimeEvolution} it is enough to diagonalize the operator at each node. The eigenvectors of the full operator are tensor product of eigenvector of operators restricted to each node. In the homogeneous-isotropic sector the situation is even simpler, because the isotropic eigenvectors of the full operator are tensor products of one and the same eigenvector corresponding to a fixed node. Furthermore, we argued in \cite{HomogeneousIsotropic} that it is enough to restrict to a subspace (see also \eqref{eq:Hilbert_space})
\begin{equation}
\Hil^{\rm loop}_{\Gamma,j,\node}=\bigoplus_{\loopconfiguration} \Hil_{\loopconfiguration}
\end{equation}
and treat the quantum Hamiltonian as acting in the direct product of the intertwiner spaces. We consider a model of gravity coupled to a massless scalar field and use the Hamiltonian in the form proposed in \cite{TimeEvolution}:
\begin{equation}
H=\sqrt{-\widehat{\sqrt{q}C^{\rm gr}}}=\sqrt{-\frac{E_{\rm PL} \ell_{\rm PL}^3}{16 \pi \gamma^2}\left((1+\gamma^2)\CL+\CE^{\dagger} + \CE\right)+\frac{\Lambda}{\kappa} \hat{V}^2}.
\end{equation}
In the formula above: $E_{\rm PL}$ is the Planck energy, $\ell_{\rm PL}$ is the Planck length, $\gamma$ is the Barbero-Immirzi parameter (which we will take to be $0.24$), $\Lambda$ is the cosmological constant and $\kappa=\frac{8\pi G}{c^4}$. The operator $\hat{V}$ is the volume operator which preserves each space $\Hil_\loopconfiguration$ and in fact is proportional to the identity operator when restricted to each $\Hil_\loopconfiguration$ (see \cite{VolumeMonochromatic} for proof). $\CL$ is a quantum operator corresponding to classical term $\int d^3 x (\sqrt{q})^2 R^{(3)}$, where $q$ is the spatial (3d) metric and $R^{(3)}$ is the corresponding Ricci scalar. In the proposal \cite{HamiltonianOperator,NewScalarConstraint,TimeEvolution} this operator does not change the graph -- in our notation this means that $\CL \Hil_\loopconfiguration \subset \Hil_\loopconfiguration$. We will not use this operator in this paper, because as we argue in section \ref{sc:internal_flatness} the internal geometries are flat and we expect that with proper definition this part (called in \cite{HamiltonianOperator,NewScalarConstraint,TimeEvolution} Lorentzian part) vanishes on our states or at least is negligibly small compared to the Euclidean part and cosmological constant term.

The nontrivial operator is the Euclidean part: $\CE$ and $\CE^\dagger$. The operator $\CE^{\dagger}$ adds loops and the operator $\CE$ is its hermitian adjoint, in particular it subtracts loops. The operator $\CE^\dagger$ is further defined by a sum of operators
\be
\CE^{\dagger}=\sum_{r,s} \epsilon(\dot{\link}_r\,\dot{\link}_s) \CE^\dagger_{rs},
\ee
where $\epsilon(\dot{\link}_r\,\dot{\link}_s)$ was defined in \eqref{eq:epsilon} and the sum runs over all possible pairs of links $(\link_r,\link_s)$ at the at least $6$-valent node. The operator $\CE^\dagger_{rs}$ maps elements of $\Hil_{\loopconfiguration}$ into elements of $\Hil_{\loopconfiguration_{rs}}$, where $\numberofloops(\loopconfiguration_{rs})=\numberofloops(\loopconfiguration)+1$. Let us underline that with $I,J,R,S=1,\ldots,6$ we label the sides of the cubical lattice and with $r,s=1,\ldots, 2\numberofloops(\loopconfiguration)+6$ we label the links outgoing from the fixed node $\node$. The loop configuration $\loopconfiguration_{rs}$ is constructed in the following way. Let $\link_R$ be a side of the cubical lattice to which $\link_r$ is tangent and $\link_S$ be a side of the cubical lattice to which $\link_s$ is tangent (a link is tangent to itself). The loop configuration $\loopconfiguration_{rs}$ has the following form:
\begin{equation}
\loopconfiguration_{rs}(IJ)=\begin{cases}
\loopconfiguration(I,J)+1,&\textrm{ if }(I,J)=(R,S)\textrm{ or }(I,J)=(S,R),\\
\loopconfiguration(I,J)&\textrm{otherwise}.
\end{cases}
\end{equation}
The action of the operator $\CE_{rs}^{\dagger}$ on the intertwiner spaces is given in the next subsection.

\subsection{Definition of the Euclidean part}
Consider a sequence of representations $\rho_{j_1},\ldots, \rho_{j_N}$ and a sequence of their representation spaces $\Hil_{j_1}, \ldots, \Hil_{j_N}$. In the space $\Hil_{j_1}\otimes \ldots \otimes \Hil_{j_N}$ we define operators $\hat{J}_{r\,i}, r\in\{1,\ldots, N\}, i\in\{1,2,3\}$ by the following formula:
\be
\hat{J}_{r\,i}:=\id \otimes \id\otimes \rho'_r(\tau_i)\otimes \id \otimes\ldots \otimes \id,
\ee
where $\tau_i=-\frac{\iu}{2}\sigma_i$ form the su(2) Lie algebra basis defined by the Pauli matrices $\sigma_i$ and  $\rho'_r$ is the su(2) representation corresponding to $\rho_r$. 

As we discussed in detail in \cite{SFscalar}, the Euclidean part of the Hamiltonian operator is defined by a family of operators 
\be
\CE^\dagger_{rs}:\Hil_{j_1}\otimes\ldots \otimes \Hil_{j_N}\to \Hil_{j_1}\otimes\ldots\otimes \Hil_{j_r}\otimes\Hil_{1/2} \otimes\ldots \otimes \Hil_{1/2}^*\otimes \Hil_{j_s}\otimes \ldots \otimes \Hil_{j_N}
\ee
such that (repeated indices are summed over)
\be
\CE^\dagger_{rs}=8 \kappa_1 \epsilon_{ijk} \tau_i \hat{J}^j_r \hat{J}^k_s.
\ee
In this formula we treat $\tau_i$ as a tensor in $\Hil_{1/2}\otimes  \Hil_{1/2}^*$. The coefficient $\kappa_1$ is a real number which value can depend only on the valence of the intertwiners \cite{HamiltonianOperator}. We will denote by $\CE_{rs}$ the hermitian adjoint of $\CE^\dagger_{rs}$. 

In this paper we split each loop created by the Euclidean part into 2 links by adding a 2-valent node $\node$ and re-orienting the links such that all links are outgoing from the node which is at least $6$-valent. Therefore we will consider an equivalent family of operators (which will be denoted by the same symbol):
\be
\CE^\dagger_{rs}:\Hil_{j_1,\ldots,j_N} \to \Hil_{j_1,\ldots,j_r,1/2,\ldots,1/2,j_s,\ldots,j_N}\otimes \left(\Hil_{1/2}^* \otimes \Hil_{1/2}^* \right).
\ee
In the formula above, the Hilbert spaces $\Hil_{j_1,\ldots,j_N}$ and $\Hil_{j_1,\ldots,j_r,1/2,\ldots,1/2,j_s,\ldots,j_N}$ correspond to the at least $6$-valent node $\node_0$ and the Hilbert space $\Hil_{1/2}^* \otimes \Hil_{1/2}^*$ corresponds to the bivalent node $\node$. The operator $\CE^\dagger_{rs}$ becomes:
\be
\CE^\dagger_{rs}=8 \kappa_1 \epsilon_{ijk} \tilde{\tau}_i \hat{J}^j_r \hat{J}^k_s\otimes \epsilon.
\ee
In the formula above $\epsilon:\Hil_{1/2}\to \Hil_{1/2}^*$ is the canonical intertwiner between the spin $1/2$ representation and its dual, and $\tilde{\tau}_i=\left(\id\otimes \epsilon^{-1}\right) \tau_i$. In the index notation
\be
\epsilon_{AB}=\begin{pmatrix} 0&1\\-1&0 \end{pmatrix},\quad(\epsilon^{-1})^{AB}=\begin{pmatrix} 0&-1\\1&0 \end{pmatrix}, \quad {\tilde{\tau}_i}^{A\,B}=(\epsilon^{-1})^{B\,C}{\tau_i}^A_C.
\ee

\subsection{Contributions for parallel links}
Let us denote by $\link_I, I\in\{1,\ldots,6\}$ the segments of the lattice (without loops) intersecting at the node $\node_0$. Let us introduce operators
\be
\CE^{\dagger}_{IJ}=8 \kappa_1 \epsilon_{ijk} \tilde{\tau}_i \left( \sum_{r:\dot{\link}_r \parallel \dot{\link}_I}\hat{J}^j_r \right) \left(\sum_{s:\dot{\link}_s \parallel \dot{\link}_J} \hat{J}^k_s\right)\otimes \epsilon,
\ee
where the first sum is over links $\link_r$ tangent to $\link_I$ and the second sum is over $\link_s$ tangent to link $\link_J$. The operator $\CE^{\dagger}$ can now be written in the following form:
\be
\CE^{\dagger}=\sum_{I=1}^6 \sum_{J=1}^6 \epsilon_{I\,J} \CE^\dagger_{IJ}.
\ee
The $\epsilon_{IJ}$ is $0$ if the sides of the lattice $\link_I$ and $\link_J$ are anti-parallel and $1$ otherwise. This expression is especially useful in our coherent state basis build from Perelomov coherent states. Let us recall that we assign to each link $\link_r$ tangent to $\link_I$ a coherent state $\ket{1/2,\vec{n}_I}$. We can use the standard isomorphism mapping spinors into the corresponding states in the magnetic number representation to obtain (compare \eqref{eq:spinors_iso_magnetic}):
\be
\ket{1/2, \vec{n}_I}^{\otimes N_I}= \ket{j_I,\vec{n}_I},
\ee
where $j_I=N_I/2$. Under this isomorphism we can write:
\be
\sum_{r:\dot{\link}_r \parallel \dot{\link}_I}\hat{J}^j_r \bigotimes_{r:\dot{\link}_r \parallel \dot{\link}_I}\ket{1/2, \vec{n}_r} = \hat{J}^j_I \ket{N_I/2,\vec{n}_I},
\ee
where $\hat{J}^j_I = \rho'_{j_I}(\tau^j)$. With this notation, we can write $\CE^{\dagger}_{IJ}$ in the following form:
\be\label{eq:CE_dagger}
\CE^{\dagger}_{IJ}=8 \kappa_1 \epsilon_{ijk} \tilde{\tau}_i \hat{J}^j_I\hat{J}^k_J \otimes \epsilon.
\ee
This form will be useful for calculating matrix elements between our coherent basis states. 

\subsection{Invariance of the matrix elements}
In \cite{HomogeneousIsotropic} we studied in detail the consequences of an the invariance of the Hamiltonian operator. In particular, for the operator $\CE$ we have:
\be\label{eq:CE_invariance}
\forall_{g\in\CubicalSymmetries} U_g^{\dagger}\, \CE\, U_g = \CE.
\ee
By using this property we will reduce the problem of calculating matrix elements of $\CE$ between states $\ket{k}$ into a simpler problem of calculating matrix elements of $\CE_{IJ}$ for some $I,J$ between states $\ket{\loopconfiguration_k}$. In particular, from \eqref{eq:CE_invariance} and \eqref{eq:hom_iso_states_group} it follows that 
\begin{multline}\label{eq:CEkk}
\bra{k+1} \CE^{\dagger} \ket{k}=\frac{1}{24\cdot \sqrt{\# \stabilizer{\loopconfiguration_{k+1}} \# \stabilizer{\loopconfiguration_{k}} }}\sum_{g,g'\in \CubicalSymmetries}\bra{\loopconfiguration_{k+1}} U_{g'}^\dagger \CE^{\dagger} U_g \ket{\loopconfiguration_k}=\\=\frac{1}{\sqrt{\# \stabilizer{\loopconfiguration_{k+1}} \# \stabilizer{\loopconfiguration_{k}} }}\sum_{g'\in \CubicalSymmetries}\bra{\loopconfiguration_{k+1}} U_{g'} \CE^{\dagger} \ket{\loopconfiguration_k}.
\end{multline}
Let us recall that in the formula above $\# \stabilizer{\loopconfiguration_{k}}$ is the number of elements in the stabilizer group of $\loopconfiguration_k$. We have:
\begin{align}
\# \stabilizer{\loopconfiguration_{3n}}=
\begin{cases} 
24,&{\rm if\ }n=0,\\
6,&{\rm if\ }n>0.
\end{cases},\label{eq:stabilizer_tn}\\
\# \stabilizer{\loopconfiguration_{3n+1}}=\# \stabilizer{\loopconfiguration_{3n+2}}=2.\label{eq:stabilizer_rest}
\end{align}

Let us notice that the only non-zero contribution in the sum in \eqref{eq:CEkk} comes from elements $g'$ that map links $\{\link_1,\link_3\}$ into: $\{\link_1,\link_3\}$ or $\{\link_2,\link_6\}$ or $\{\link_4,\link_5\}$. Since $g'$ is orientation preserving isometry of a cube, it is completely determined by its action on $e_x$ and $e_y$. Therefore there are $6$ possibilities: 
\begin{enumerate}
\item $g'(1)=1,g'(3)=3$,
\item $g'(1)=3,g'(3)=1$,
\item $g'(1)=2,g'(3)=6$,
\item $g'(1)=6,g'(3)=2$,
\item $g'(1)=4,g'(3)=5$,
\item $g'(1)=5,g'(3)=4$.
\end{enumerate}
Each $g'$ from the list above is a different element of $\CubicalSymmetries$. Direct calculation shows that the elements form a group -- the stabilizer of loop configuration $\loopconfiguration_{3n}, n\in \mathbb{N}_+$. Taking into account that $\CE^\dagger$ is a sum of operators $\CE^{\dagger}_{IJ}$ and that $g'$ is in the stabilizer group of $\loopconfiguration_{3n}$:
\be
\bra{\loopconfiguration_{k+1}} U_{g'} \CE^{\dagger} \ket{\loopconfiguration_k}=\sum_{I,J} \epsilon_{I\,J} \bra{\loopconfiguration_{k+1}} U_{g'} \CE^{\dagger}_{I J} \ket{\loopconfiguration_k} =2\bra{\loopconfiguration_{k+1}} U_{g'} (\CE^{\dagger}_{13}+\CE^{\dagger}_{26}+\CE^{\dagger}_{45}) \ket{\loopconfiguration_k}.
\ee
The factor of $2$ in the last equality comes from the fact that $\CE^{\dagger}_{IJ}=\CE^{\dagger}_{JI}$.

There are 3 cases:
\begin{enumerate}
\item $k=3n$. Since $g'$ is in the stabilizer group of $\loopconfiguration_k$ and $\CE^{\dagger}$ transforms according to \eqref{eq:CE_invariance}:
\be
\bra{\loopconfiguration_{k+1}} U_{g'} \CE^{\dagger} \ket{\loopconfiguration_k}=\bra{\loopconfiguration_{k+1}} \CE^{\dagger} \ket{\loopconfiguration_k}=2\bra{\loopconfiguration_{k+1}} \CE^{\dagger}_{13} \ket{\loopconfiguration_k}.
\ee
As a result:
\be
\bra{k+1} \CE^{\dagger} \ket{k}= \frac{2\cdot\# \stabilizer{\loopconfiguration_{3n}}}{\sqrt{\# \stabilizer{\loopconfiguration_{3n}}\cdot \# \stabilizer{\loopconfiguration_{3n+1}}}} \bra{\loopconfiguration_{k+1}} \CE^{\dagger}_{13} \ket{\loopconfiguration_k}.
\ee
After inserting the values from \eqref{eq:stabilizer_tn} and \eqref{eq:stabilizer_rest} we obtain:
\be\label{eq:CEdagger_I}
\bra{3n+1} \CE^{\dagger} \ket{3n}= 
\begin{cases}
4 \sqrt{3} \bra{\loopconfiguration_{k+1}} \CE^{\dagger}_{13} \ket{\loopconfiguration_k},&{\rm if\ }n=0,\\
2 \sqrt{3} \bra{\loopconfiguration_{k+1}} \CE^{\dagger}_{13} \ket{\loopconfiguration_k},&{\rm if\ }n>0.
\end{cases}
\ee
\item $k=3n+1$. In this case $\bra{\loopconfiguration_{k+1}} U_{g'} (\CE^{\dagger}_{13}) \ket{\loopconfiguration_k}=0$, because after the action of $\CE^{\dagger}_{13}$ there would be $n+2$ loops between links $\link_1, \link_3$ and it is not possible to bring such loop configuration to $\loopconfiguration_{k+1}$ by acting with an orientation preserving symmetry of a cube. Therefore,
\be
\bra{\loopconfiguration_{k+1}} U_{g'} \CE^{\dagger} \ket{\loopconfiguration_k}=2\bra{\loopconfiguration_{k+1}} U_{g'} (\CE^{\dagger}_{26}+\CE^{\dagger}_{45}) \ket{\loopconfiguration_k}.
\ee
Let us notice that the transformation $h$ such that $h(1)=3, h(3)=1$ is a symmetry of the loop configuration $\loopconfiguration_{3n+1}$. Moreover: $h(2)=4, h(6)=5$ and $h(4)=2, h(5)=6$. As a result,
\be
\bra{\loopconfiguration_{k+1}} U_{g'} \CE^{\dagger}_{45} \ket{\loopconfiguration_k}=\bra{\loopconfiguration_{k+1}} U_{g'} U_h \CE^{\dagger}_{26} U_h^\dagger \ket{\loopconfiguration_k}=\bra{\loopconfiguration_{k+1}} U_{g'\cdot h} \CE^{\dagger}_{26}\ket{\loopconfiguration_k}.
\ee
Let us notice further that there are 2 possible $g$ such that
\be
\bra{\loopconfiguration_{k+1}} U_{g} \CE^{\dagger}_{26} \ket{\loopconfiguration_k}\neq 0.
\ee
First is $g=id$, second is $g$ such that $g(4)=5, g(5)=4$. They both are symmetries of $\loopconfiguration_{k+1}$. Therefore,
\be
\bra{\loopconfiguration_{k+1}} U_{g} \CE^{\dagger}_{26} \ket{\loopconfiguration_k}\neq 0 \implies \bra{\loopconfiguration_{k+1}} U_{g} \CE^{\dagger}_{26} \ket{\loopconfiguration_k}=\bra{\loopconfiguration_{k+1}} \CE^{\dagger}_{26} \ket{\loopconfiguration_k}.
\ee
As a result,
\begin{multline}\label{eq:CEdagger_II}
\bra{3n+2} \CE^{\dagger} \ket{3n+1}=\frac{2\cdot 2\cdot\# \stabilizer{\loopconfiguration_{3n+2}}}{\sqrt{\# \stabilizer{\loopconfiguration_{3n+2}}\cdot \# \stabilizer{\loopconfiguration_{3n+1}}}} \bra{\loopconfiguration_{3n+2}} \CE^{\dagger}_{26} \ket{\loopconfiguration_{3n+1}}=\\=4\cdot \bra{\loopconfiguration_{3n+2}} \CE^{\dagger}_{26} \ket{\loopconfiguration_{3n+1}}.
\end{multline}
\item $k=3n+2$. In this case $g'$ is in the stabilizer group of $\loopconfiguration_{k+1}$ and 
\be
\bra{\loopconfiguration_{k+1}} U_{g'} \CE^{\dagger} \ket{\loopconfiguration_k}=\bra{\loopconfiguration_{k+1}} \CE^{\dagger} \ket{\loopconfiguration_k}=2\bra{\loopconfiguration_{k+1}} \CE^{\dagger}_{45} \ket{\loopconfiguration_k}.
\ee
As a result,
\begin{multline}\label{eq:CEdagger_III}
\bra{k+1} \CE^{\dagger} \ket{k}=\frac{2\cdot\# \stabilizer{\loopconfiguration_{3n+3}}}{\sqrt{\# \stabilizer{\loopconfiguration_{3n+3}}\cdot \# \stabilizer{\loopconfiguration_{3n+2}}}} \bra{\loopconfiguration_{k+3}} \CE^{\dagger}_{45} \ket{\loopconfiguration_{3n+2}}=\\=2\sqrt{3}\bra{\loopconfiguration_{k+3}} \CE^{\dagger}_{45} \ket{\loopconfiguration_{3n+2}}.
\end{multline}
\end{enumerate}

In this paper we will use non-trivially the fact the $\kappa_1$ may depend on the valence of the intertwiners (and therefore on the number of loops $k$) \cite{HamiltonianOperator}. We will assume that $\kappa_1$ is the following function of $k$:
\begin{equation}
\kappa_1(k)=\begin{cases}
\frac{\kappa_E}{4\sqrt{3}},&\textrm{ if } k=0,\\
\frac{\kappa_E}{2\sqrt{3}},&\textrm{ if } k=3n+2 \textrm{ or }k=3n+3,\\
\frac{\kappa_E}{4},&\textrm{ if } k=3n+1,\\
\end{cases}
\end{equation}
where $\kappa_E$ does not depend on $k$ and $n\in\mathbb{N}$.

\subsection{Matrix elements of the euclidean part between our coherent states}\label{sc:euclidean_asymptotic}
We will calculate now explicitly the matrix elements on the right hand side of equations \eqref{eq:CEdagger_I},\eqref{eq:CEdagger_II},\eqref{eq:CEdagger_III}. Let us notice that they are given by matrix elements of an SU(2) invariant operator between Livine-Speziale coherent intertwiners. Each of the Livine-Speziale coherent intertwiner (in and out) is obtained by averaging a tensor product of Perelomov coherent states over the SU(2) group. Due to the invariance of the operator, we can omit one such averaging. This allows us to write the expression for matrix elements in the following form:
\be\label{eq:CE_IJ}
\bra{k+1} \CE^{\dagger}_{IJ} \ket{k}=8\kappa_E \sum_{i,j,l} \epsilon_{ijl} \frac{\alpha_{IJ}^i}{N_{k}N_{k+1}}\cdot \beta^{jl}_{IJ}
\ee
In the formula above, $\alpha_{IJ}^i$ is combining two factors: a factor coming from a contraction of indices of $\tilde{\tau}_i$ with Perelomov coherent states $\ket{1/2,\vec{n}_I}, \ket{1/2,\vec{n}_J}$ in the out state and a factor coming from the contraction of $\epsilon$ assigned to bivalent node in \eqref{eq:CE_dagger} with the Livine-Speziale coherent intertwiner assigned to the bivalent node (see section \ref{sc:coherent_intertwiners}):
\be
\alpha_{IJ}^i= 2\ \epsilon(\vec{n}_I,\vec{n}_J)\cdot \left(\left(\ket{1/2,\vec{n}_I}\otimes \ket{1/2,\vec{n}_J}\right) \lrcorner \tilde{\tau}_i\right).
\ee
The factor $N_k$ is the normalization factor of the Livine-Speziale coherent intertwiners (see also \eqref{eq:Nk}):
\be
N_k = \sqrt{ \braket{\loopconfiguration_{k}, \vec{n}_1\ldots \vec{n}_6}{\loopconfiguration_{k}, \vec{n}_1\ldots \vec{n}_6}  }.
\ee
The remaining factor is $\beta^{jk}_{IJ}$ (see also \eqref{eq:beta_jl_IJ}):
\begin{multline}
\beta^{jl}_{IJ}=\int_{\rm SU(2)} du \bra{j_I,\vec{n}_I}\hat{J}_I^j\, \rho_{j_I}(u) \ket{j_I,\vec{n}_I} \bra{j_J,\vec{n}_J}\hat{J}_J^l\, \rho_{j_J}(u) \ket{j_J,\vec{n}_J}\cdot \\ \cdot \prod_{K\not\in \{I,J\}}\bra{j_K,\vec{n}_K}\rho_{j_K}(u) \ket{j_K,\vec{n}_K}.
\end{multline}

Let us calculate first $\alpha^i_{IJ}$. In the index notation:
\begin{multline}
\alpha^i_{IJ}=2\, \vec{n}_I^A \epsilon_{AB} \vec{n}_J^B\cdot \overline{\vec{n}}\indices{_I_A} \tau\indices{_i^A_C} \epsilon\indices{^C^B} \overline{\vec{n}}\indices{_J_B}=-2\, \vec{n}_J^A \epsilon_{AB} \vec{n}_I^B\cdot \overline{\vec{n}}\indices{_I_A} \tau\indices{_i^A_C} \epsilon\indices{^C^B} \overline{\vec{n}}\indices{_J_B}=\\=\frac{\iu}{4} \Tr{\epsilon(\id+\vec{n}_J\cdot \vec{\sigma})^T \epsilon (\id+\vec{n}_I\cdot \vec{\sigma})\sigma_i}=\frac{\iu}{4} \Tr{(-\id+\vec{n}_J\cdot \vec{\sigma})  (\id+\vec{n}_I\cdot \vec{\sigma})\sigma_i}=\\=\frac{1}{2}\iu(n_J^i-n_I^i)-\frac{1}{2}\epsilon_{jki} n_J^j n_I^k.
\end{multline}

Let us recall that in section \ref{sc:flatness_large_volume} we argued that the large volume limit translates in our model into a limit of large number of loops $k$. We can directly apply the results of the previous section to calculate the factor $\frac{\beta^{jl}_{IJ}}{N_k N_{k+1}}$ in this limit:
\begin{equation}
\frac{\beta^{jl}_{IJ}}{N_k N_{k+1}}\approx -\frac{(n+1)^2}{16} n^j_I n^l_J.
\end{equation} 

It is now straightforward to calculate $\bra{k+1} \CE^{\dagger}_{IJ} \ket{k}$ in the limit of large volumes:
\begin{multline}
\bra{k+1} \CE^{\dagger}_{IJ} \ket{k}\approx- 8\kappa_E\sum_{i,j,l}\epsilon_{ijl} \left( \frac{1}{2}\iu(n_J^i-n_I^i)-\frac{1}{2}\epsilon_{j'k'i} n_J^{j'} n_I^{k'}\right) \frac{(n+1)^2}{16} n^j_I n^l_J=-\kappa_E\frac{(n+1)^2}{4}.\label{eq:asymptotic_CE}
\end{multline}
\subsection{Numerical analysis}\label{sc:euclidean_numerical}
%\begin{figure}[!tbp]
%\centering
%\subfloat[$k=3n$]{\input{CE_I.tex}}\label{fig:CE_I}
%\hfill
%\subfloat[$k=3n+1$]{\input{CE_II.tex}}\label{fig:CE_II}
%\hfill
%\subfloat[$k=3n+2$]{\input{CE_III.tex}}\label{fig:CE_III}
%\caption{Matrix elements $\frac{1}{\kappa_E}\bra{k+1} \CE^{\dagger} \ket{k}$ as a function of $n$. We plot separately the 3 cases $k=3n$, $k=3n+1$, $k=3n+2$ and fit a quadratic function $f(x)=a\,x^2 +b\,x +c$ in each case separately. The fitted function is plotted with a dotted line. }\label{fig:CE_matrix_elements}
%\end{figure}

\begin{figure}[!tbp]
\centering
\includegraphics{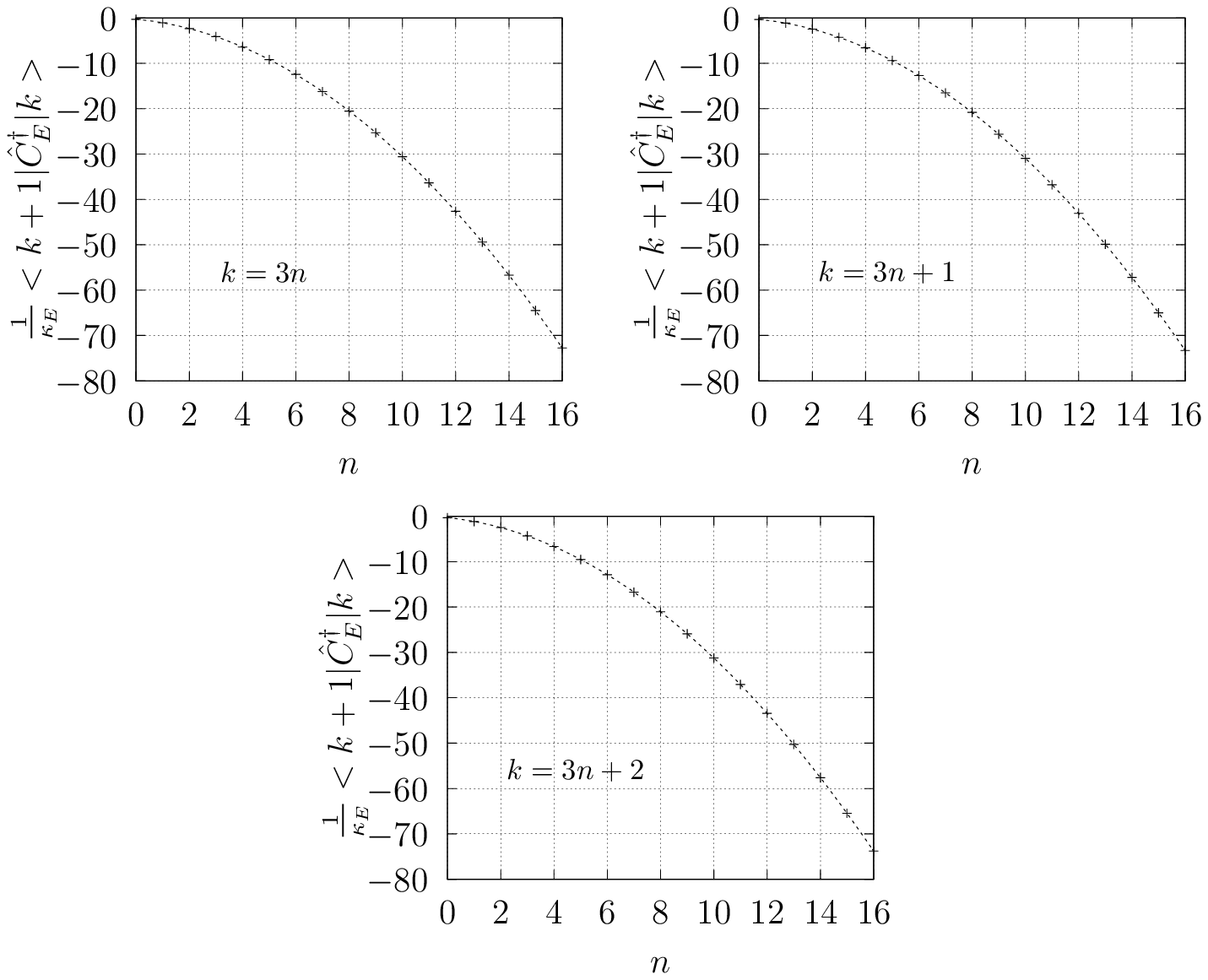}
\caption{Matrix elements $\frac{1}{\kappa_E}\bra{k+1} \CE^{\dagger} \ket{k}$ as a function of $n$. We plotted the 3 cases $k=3n$, $k=3n+1$, $k=3n+2$ and fitted a quadratic function $\frac{1}{\kappa_E}\bra{k+1} \CE^{\dagger} \ket{k}=a\,n^2 +b\,n +c$ in each case separately. The fitted function is plotted with a dotted line. }\label{fig:CE_matrix_elements}
\end{figure}

We have calculated the matrix elements $\frac{1}{\kappa_E}\bra{k+1} \CE^{\dagger} \ket{k}$ numerically for $k\in \{0,1,\ldots,50\}$. The results are summarized in figure \ref{fig:CE_matrix_elements}. We investigated the dependence of $\frac{1}{\kappa_E}\bra{k+1} \CE^{\dagger} \ket{k}$ as a function of $n$. Using gnuplot software we fitted a quadratic function $\frac{1}{\kappa_E}\bra{k+1} \CE^{\dagger} \ket{k}=a\,n^2 +b\,n +c$ in 3 cases $k=3n$, $k=3n+1$, $k=3n+2$ separately (see figure \ref{fig:CE_matrix_elements}). We obtained the following results
\begin{eqnarray*}
a_{3n}= -0.250164 \pm 0.000034,\quad b_{3n}= -0.527623 \pm 0.0005636,\quad c_{3n}=-0.275961 \pm 0.001944,\\
a_{3n+1}= -0.249966\pm 0.000017,\quad b_{3n+1}= -0.563023 \pm 0.0002825,\quad c_{3n+1}=-0.321144\pm 0.0009746,\\
a_{3n+2}= -0.249991 \pm 0.000002,\quad b_{3n+2}= -0.593974 \pm 0.000031,\quad c_{3n+2}=-0.308724 \pm 0.0001085.
\end{eqnarray*}
This confirms our asymptotic result \eqref{eq:asymptotic_CE} predicting that in the limit of large $n$ :
\be
\bra{k+1} \CE^{\dagger} \ket{k}\approx-\frac{(n+1)^2}{4}\kappa_E.
\ee The coefficient of the quadratic term in the fitted functions are in good accordance with the asymptotic value $-0.25$. The appearance of the linear term is expected, because our formula \eqref{eq:asymptotic_CE} holds in the limit of large $n$ only. We expect that the values of the linear coefficients can be calculated by considering subleading order contributions in the saddle point analysis. In this paper, we determine them from our numerical experiment. Strictly speaking, we approximate the matrix elements (for any $n$) with
\be\label{eq:CE_approximate}
\frac{1}{\kappa_E}\bra{k+1} \CE^{\dagger} \ket{k}\approx -\frac{(n+1)^2}{4}- B n - C.
\ee
The coefficients $B$ and $C$ are obtained by fitting the function on the right hand side to our numerical data in the 3 cases $k=3n$, $k=3n+1$, $k=3n+2$ separately. We obtained:
\begin{eqnarray*}
B_{3n}= 0.0302524 \pm 0.0002339,\quad C_{3n}=0.0193867 \pm 0.002194,\\
B_{3n+1}= 0.0624754\pm 0.0000814,\quad C_{3n+1}= 0.0725132\pm 0.0007639,\\
B_{3n+2}= 0.0938279 \pm 0.000013,\quad C_{3n+2}= 0.0590887\pm 0.0001219.
\end{eqnarray*}
We calculated the matrix elements by a straightforward implementation of formula \eqref{eq:CE_IJ}. In order to calculate $\beta^{jl}_{IJ}$ and $N_k$ we constructed a tree basis of intertwiners by contracting 3j-symbols. Using this tree basis we constructed a projection onto invariant tensors. Next, we projected   a tensor product of Perelomov coherent states and obtained the Livine-Speziale coherent intertwiners. The implementation of angular momentum operators was straightforward as it is a standard textbook formula. The coefficients $\beta^{jl}_{IJ}$ and $N_k$ were obtained by further contractions. In our algorithm the number of contractions grows very fast with increasing the number of loops. As a result we were able to calculate the matrix elements up to $50$ loops (it took around 1 week on 1 node equipped with 40 CPUs and 128 GB RAM). While we expect that it is possible to derive a computationally faster method, it will be enough for us to use the approximate formula \eqref{eq:CE_approximate}.

\section{Cosmological model derived from the full Loop Quantum Gravity}

\subsection{Cosmological Hamiltonian}
In our model we consider a Hilbert space $\Hil_{\rm flat}$ spanned by states $\ket{k}$. It is a subspace of homogenous-isotropic states proposed in \cite{HomogeneousIsotropic} which we interpret as a space of quantum geometries with zero intrinsic curvature. Indeed, in section \ref{sc:internal_flatness} we discuss the internal geometry of the states and conclude that the geometries are approximately flat. Therefore the quantum operator corresponding to the term $\int d^3x \sqrt{q}^2 R^{(3)}$ should anihilate the states or at least it's expectation values in such states should be negligibly small compared to the expectation values of the Euclidean part and the cosmological constant term (in the limit of large volumes). The gravitational part of the scalar constraint operator $\widehat{\sqrt{q} C^{\rm gr}}$ restricted to this space has only diagonal and subdiagonal entries. The diagonal entries are given by the cosmological constant term (in the case of zero intrinsic curvature):
\be
\bra{k}{\widehat{\sqrt{q} C^{\rm gr}}}\ket{k}\approx-\frac{\Lambda}{\kappa} \bra{k}\hat{V}^2\ket{k}=-\left(\Lambda l_P^2 \right) \frac{ \sqrt{3}  \pi^2 \gamma^3 \kappa_0^2}{6} (N-2)N(N+2) E_P l_P^3,
\ee
where $N=2k+6$. In the formula above $\gamma$ is the Barbero-Immirzi parameter (which we will assume to be $\gamma=0.24$), $\kappa_0$ is a free constant in the definition of the volume operator. The subdiagonal elements are $\bra{k+1} \CE^\dagger \ket{k}$ which we approximate according to formula \eqref{eq:CE_approximate}:
\be
\bra{k+1}{\widehat{\sqrt{q} C^{\rm gr}}}\ket{k}=-\frac{E_P l_P^3 \kappa_E}{16 \pi \gamma^2} \left(\frac{(n+1)^2}{4}+ B_k n + C_k\right).
\ee
The operator $\sqrt{-\sqrt{q} C^{\rm gr}}$ plays a role of true Hamiltonian in a deparametrized model of Loop Quantum Gravity coupled with a massless scalar field \cite{GravityQuantized}. The solutions to the constraints are of the following form:
\be
\exp(\frac{\iu}{\hbar c}\sum_{x\in \Sigma} \sqrt{-\widehat{\sqrt{q} C^{\rm gr}}_x} \hat{\phi}_x )\Psi(A),
\ee
where $\Psi(A)$ is a solution to the Gauss and diffeomorphism constraints \cite{GravityQuantized}. In this paper we will consider a simplified model which evolution is goverened by Hamiltonian 
\be \label{eq:cosmological_hamitlonian}
\hat{H}=\sqrt{-\hat{\Theta}_{\lambda}},
\ee where $\hat{\Theta}_\Lambda$ is a projection of $\sqrt{-\sqrt{q} C^{\rm gr}}$ onto  $\Hil_{\rm flat}$, i.e. it has only diagonal and subdiagonal components:
\begin{eqnarray}
\bra{k}{\hat{\Theta}_\Lambda}\ket{k}=- E_P l_P^3 \left(\Lambda l_P^2 \right) \frac{4 \sqrt{3}  \pi^2 \gamma^3 \kappa_0^2}{3} (k+2)(k+3)(k+4),\\
\bra{k}{\hat{\Theta}_\Lambda}\ket{k+1}=\bra{k+1}{\hat{\Theta}_\Lambda}\ket{k}=-\frac{E_P l_P^3 \kappa_E}{16 \pi \gamma^2} \left(\frac{(\lfloor \frac{k}{3}\rfloor+1)^2}{4}+ B_k \lfloor \frac{k}{3}\rfloor + C_k\right).
\end{eqnarray}
\subsection{Semiclassical limit}\label{sc:semiclassical}
Let $\psi(x)$ be a real valued, square integrable function on $\mathbb{R}_+$. We consider a state $\ket{\psi}$ in $\Hil_{\rm flat}$ defined by
\be
\ket{\psi}=\sum_k \psi(k) \ket{k}.
\ee
We will assume that $\psi(x)=0$ for $x<x_0$, where $x_0$ is sufficiently large (it is actually enough to consider $\psi$ which is sufficiently small in this region). As a result $\ket{\psi}$ has non-zero components in the region where $k$ is large, so it is non-zero in the region where the volume of the Universe is sufficiently large. This is typically considered to be a semiclassical limit of a quantum cosmological model \cite{BRV}. Let us consider an action of the operator $\hat{\Theta}_\Lambda$ on this state:
\be
- \hat{\Theta}_\Lambda\ket{\psi}=\sum_k f_+(k) \psi(k-1)\ket{k} + \sum_k f_0(k) \psi(k)\ket{k} + \sum_k f_-(k) \psi(k+1)\ket{k},
\ee
where
\begin{align}
f_+(k+1)=-\bra{k+1}\hat{\Theta}_\Lambda\ket{k}=\frac{E_P l_P^3 \kappa_E}{16 \pi \gamma^2} \left(\frac{(\lfloor \frac{k}{3}\rfloor+1)^2}{4}+ B_k \lfloor \frac{k}{3}\rfloor + C_k\right),\\
f_-(k-1)=-\bra{k-1}\hat{\Theta}_\Lambda\ket{k}=f_+(k),\\
f_0(k)=-\bra{k}\hat{\Theta}_\Lambda\ket{k}=E_P l_P^3 \left(\Lambda l_P^2 \right) \frac{4 \sqrt{3}  \pi^2 \gamma^3 \kappa_0^2}{3} (k+2)(k+3)(k+4)
\end{align}
Let us introduce a function 
\be
g(k):=\frac{1}{4}\,f_+(k+\frac{1}{2}).
\ee
Using this function we can write the action of $- \hat{\Theta}_\Lambda$ on the function $\Psi$ in the following form:
\be\label{eq:theta_finite_step}
(-\hat{\Theta}_\Lambda \psi)(k)=4\, g(k+\frac{1}{2})\,\psi(k+1)+4\, g(k-\frac{1}{2})\,\psi(k-1)-4\,\left(g(k+\frac{1}{2})+g(k-\frac{1}{2})\right)\psi(k)+\tilde{f}_0(k)\psi(k),
\ee
where
\be\label{eq:modified_cosmological_term}
\tilde{f}_0(k)=E_P l_P^3 \left(\Lambda l_P^2 \right) \frac{4 \sqrt{3}  \pi^2 \gamma^3 \kappa_0^2}{3} (k+2)(k+3)(k+4)+4\,\left(g(k+\frac{1}{2})+g(k-\frac{1}{2})\right)\psi(k).
\ee
For sufficiently large $k$ the third order term in $\tilde{f}_0(k)$ dominates over the quadratic, linear and constant terms and we can approximate $\tilde{f}_0(k)$ with $f_0(k)$:
\be\label{eq:modified_cosmological_term_approximation}
\tilde{f}_0(k)\approx f_0(k).
\ee
 Additionally, in this limit we can approximate the first three terms in \eqref{eq:theta_finite_step} with a second order derivative:
\be
(-\hat{\Theta}_\Lambda \psi)(k)\approx \partial_k \left(g(k)\partial_k\psi(k)\right)+f_0(k)\psi(k).
\ee
In this limit we will approximate $g(k)$ with the quadratic term only
\be
g(k)=\frac{E_P l_P^3 \kappa_E}{16\cdot 9 \pi \gamma^2} k^2.
\ee
As a result we obtain:
\be
(-\hat{\Theta}_\Lambda \psi)(k)\approx \frac{E_P l_P^3 \kappa_E}{16\cdot 9 \pi \gamma^2}\partial_k \left(k^2\partial_k\psi(k)\right)+f_0(k)\psi(k).
\ee
Next, we will assume that $k\partial_k (k\partial_k \psi) \gg  k\partial_k \psi $. This assumption can be justified a posteriori, but similar assumption is made in Loop Quantum Cosmology \cite{AshtekarPawlowskiCosmologicalConstant}. Under this assumption we will write:
\be
(-\hat{\Theta}_\Lambda \psi)(k)\approx \frac{E_P l_P^3 \kappa_E}{16\cdot 9 \pi \gamma^2}k\partial_k \left(k\partial_k\psi(k)\right)+f_0(k)\psi(k).
\ee
Let us recall that the states $\ket{k}$ are eigenvectors of the volume operator:
\be
\hat{V} \ket{k}=\frac{\kappa_0}{8} \left(\frac{8\pi G \hbar \gamma}{c^3}\right)^{\frac{3}{2}} \sqrt{\frac{4\sqrt{3}}{3} (k+2)(k+3)(k+4)} \ket{k}.
\ee
We can therefore change to volume eigenbasis easily. Let us consider a change of variables:
\be
v(k)=\frac{\kappa_0}{8} \left(\frac{8\pi G \hbar \gamma}{c^3}\right)^{\frac{3}{2}} \sqrt{\frac{4\sqrt{3}}{3} (k+2)(k+3)(k+4)}.
\ee
In the large $k$ limit we can approximate $v(k)$ :
\be
v(k)\approx\frac{\kappa_0}{8} \left(\frac{8\pi G \hbar \gamma}{c^3}\right)^{\frac{3}{2}} \sqrt{\frac{4\sqrt{3}}{3}}\, k^{\frac{3}{2}}.
\ee
Let us notice that
\be
k\partial_k= \frac{3}{2} v\partial_v.
\ee
We obtain:
\be
-\hat{\Theta}_\Lambda \psi\approx \frac{E_P l_P^3 \kappa_E}{64 \pi \gamma^2}v\partial_v \left(v\partial_v\psi\right)+\frac{\Lambda}{\kappa} v^2\psi.
\ee
We notice that this result coincides with the Wheeler-deWitt equation if we set
\be
\kappa_E=12\pi\cdot 64\pi \gamma^2=3\cdot(16\pi\gamma)^2.
\ee

The semiclassical approximation is valid for sufficiently large $k$. Let us underline that the region where this approximation holds depends on the cosmological constant $\Lambda$. In equation \eqref{eq:modified_cosmological_term} we assumed that the third order term dominates over the second order term. For small (but positive) $\Lambda$ this may require taking very large $k$. On the other hand, the diagonal terms of the $\Theta_\Lambda$ matrix are third order in $k$ but the offdiagonal terms are only second order. This means that for sufficiently large $k$ the matrix becomes approximately diagonal, the off-diagonal part is only a small perturbation. This means that eigenvectors are localized in a volume region. As we will see shortly, this will define an upper bound for possible volumes. As a result the semiclassical approximation is valid only in a region where $k$ is large enough (the approximation \eqref{eq:modified_cosmological_term_approximation} should hold ) but not big enough  ( the offdiagonal terms cannot be neglected when compared to the diagonal terms). We will notice that in the region of small $k$ the quantum gravity effects will resolve the singularity and lead to a Big Bounce. On the other hand for very large $k$ the quantum gravity effects will cause a Big Crunch.

\subsection{Eigenvectors and eigenvalues of the cosmological Hamiltonian}\label{sc:spectral_properties}
\begin{figure}[!tbp]
\centering
\includegraphics{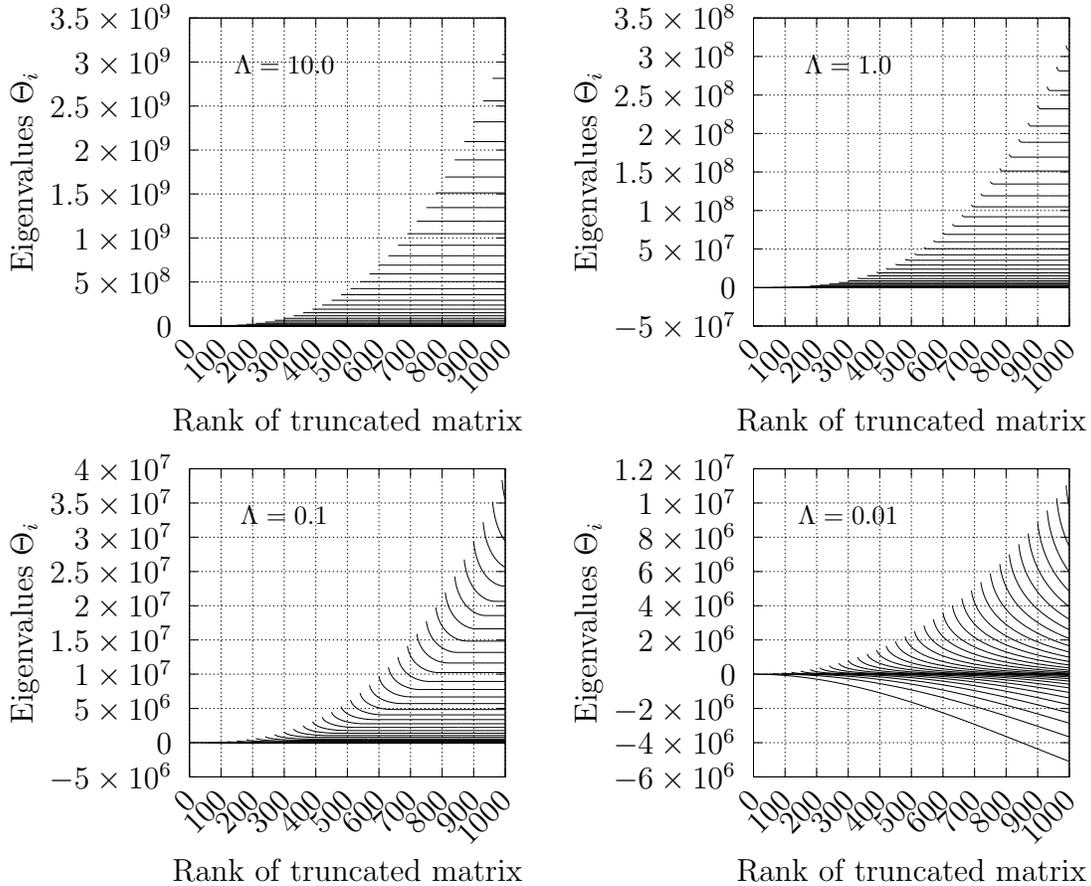}
\caption{The plot illustrates a convergence of eigenvalues of $\hat{\Theta}_\Lambda$ when the cut-off is increased. Each plot corresponds to different value of the cosmological constant $\Lambda$ (in Planck units). We plotted some of the eigenvalues of the matrices obtained by truncating $\hat{\Theta}_\Lambda$ to spaces with up to $0,1,2,\ldots, 999$ loops. The eigenvalues were sorted in increasing order. Every 30th eigenvalue is plotted starting with the lowest eigenvalue, i.e. we plot the eigenvalues $\Theta_1,\Theta_{31},\Theta_{61},\ldots, \Theta_{991}$ as functions of the truncation. In order to guide the eye we connected with a line each i-th eigenvalue in each truncation. Let us notice that $\Theta_i$ appears in matrices of rank at least $i$.}\label{fig:Theta_eigenvalues}
\end{figure}

\begin{figure}[!tbp]
\centering
\includegraphics{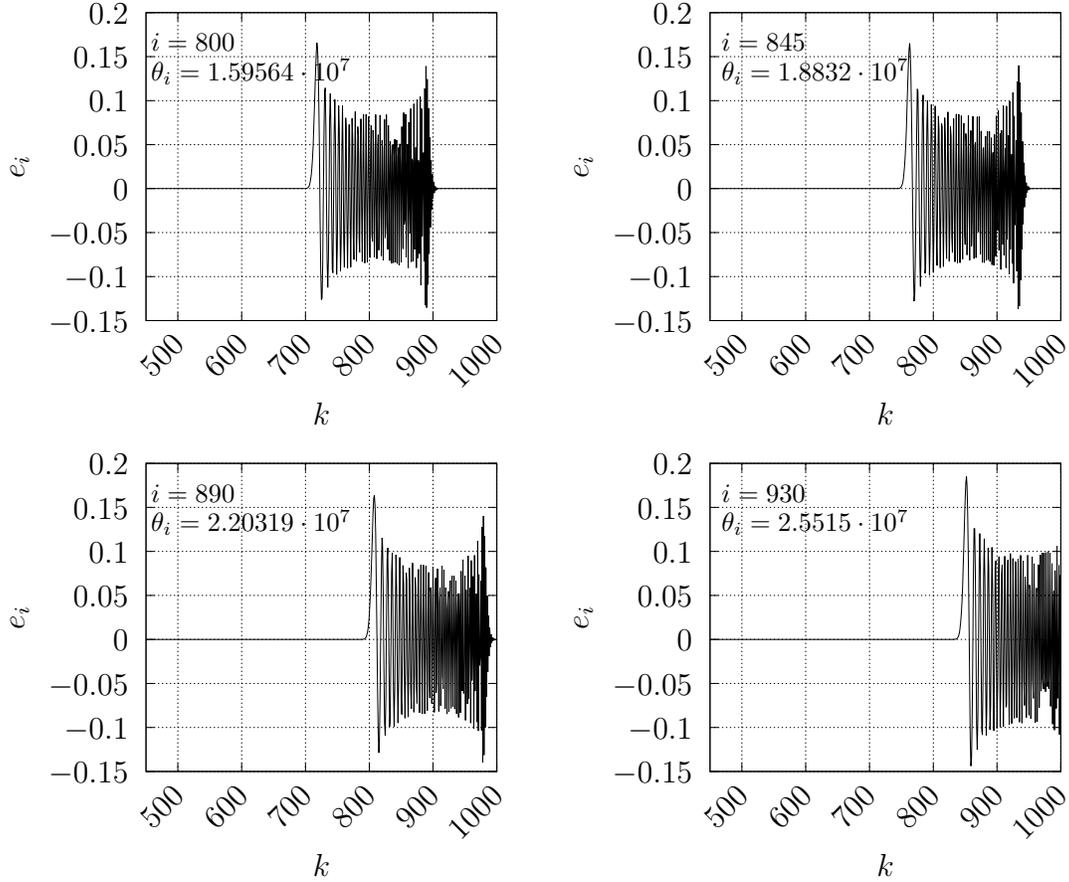}
\caption{The plot explains the convergence properties of the eigenvalues (and eigenvectors). The case of $\Lambda=0.1$ is depicted. We chose eigenvectors corresponding to eigenvalues $\Theta_i$, where $i\in\{800,845,890,930\}$. The eigenvectors are localized in a volume range. If the range is outside of the truncation range, the truncated eigenvalues and eigenvectors do not approximate well the full eigenvectors. }\label{fig:Theta_eigenvectors_convergence}
\end{figure}

\begin{figure}[!tbp]
\centering
\includegraphics{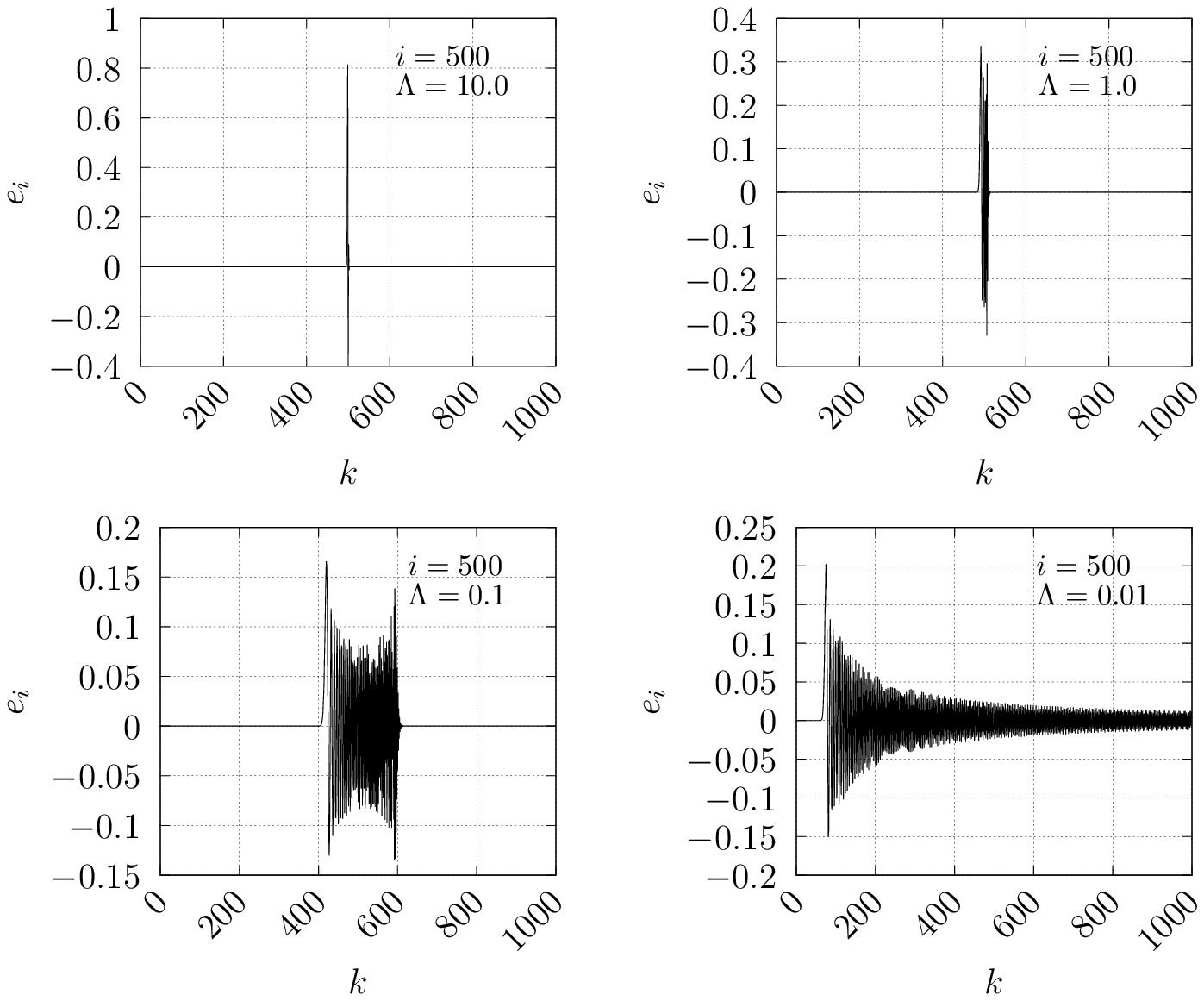}
\caption{The plot illustrates the dependence of the eigenvectors $e_i$ on the cosmological constant $\Lambda$. The range where the eigenvector is localized widens as we decrease $\Lambda$ (still we assume that $\Lambda>0$). In the last plot ($\Lambda=0.01$) the range is outside of the truncation region. This reflects the fact that a larger matrix is needed to obtain good approximation of an eigenvector of the full (non-truncated) operator. }\label{fig:Theta_eigenvectors_various_Lambda}
\end{figure}

\textbf{In the following, we will use Planck units $\hbar=c=G=1$}.

In this section we will study spectral properties of $\hat{\Theta}_\lambda$ numerically. We will introduce a cut-off in the volume (or equivalently in the number of loops) and diagonalize the resulting (finite dimensional) matrix. The resulting eigenstates will vanish exponentially outside a finite interval $[v_1,v_2]$ in the volume space. This will allow us to conclude that the eigenvalues obtained numerically are a good approximation of the full eigenvalues if the rank of the matrix is big enough to include the full range $[v_1,v_2]$.

We start with studying the convergence of the eigenvalues. Let us sort the eigenvalues in the increasing order and denote by $\Theta_i$ the $i$-th eigenvalue. We will consider matrices obtained by truncating $\hat{\Theta}_\Lambda$ to spaces with up to $0,1,2,\ldots, 999$ loops. On figure \ref{fig:Theta_eigenvalues} eigenvalues $\Theta_1,\Theta_{31}, \Theta_{61},\ldots,\Theta_{991}$ are plotted as functions of the truncation (let us underline that $\Theta_i$ appears in matrices truncated to $i-1$ or more loops). We will say that a numerical eigenvalue $i$ from the plot \ref{fig:Theta_eigenvalues} converges if there is a plateau on the plot of $\Theta_i$ as a function of the intermediate truncations. We notice that as $\Lambda$ decreases, less and less eigenvalues converge. As a result, the smaller $\Lambda$ is, the bigger matrix needs to be considered to find eigenvalues of the full (non-truncated) operator. This property can be understood by studying the properties of the eigenvectors (see figures \ref{fig:Theta_eigenvectors_various_Lambda} and \ref{fig:Theta_eigenvectors_convergence}). Each eigenvector $e_i$ is localized in a range of number of loops (or equivalently a volume range), i.e.  vanishes for $k$ smaller than some value $k_{\rm min}(i)$ and  for $k$ bigger that some value $k_{\rm max}(i)$. An eigenvalue $e_i$ of a truncated matrix approximates well an eigenvalue of the non-truncated operator when $k_{\rm max}(i)$ is smaller than the rank of the truncated matrix. Let us notice that the smaller $\Lambda$ is, the bigger the middle eigenvalue on figure \ref{fig:Theta_eigenvectors_various_Lambda} is. This explains why for $\Lambda=0.01$ we cannot obtain reliable approximation of the full eigenvalues by considering a matrix truncated up to $999$ loops. Furthermore, let us notice that on figure \ref{fig:Theta_eigenvectors_convergence} the eigenvectors corresponding to higher eigenvalues are shifted more to the right. This explains why the eigenvectors with the highest eigenvalues cannot be approximated reliably with the eigenvectors of the truncated matrix (compare for example with the case $\Lambda=0.1$ on figure \ref{fig:Theta_eigenvalues} ).

The fact that each eigenvalue is localized in a range of number of loops has improtant physical consequences. As noted in \cite{ZhangBouncingUniverse} the fact there exists $k_{\rm min}(i)$ is a signal of a Big Bounce. In our model we noticed that each eigenvector $e_i$ also vanishes for $k$ bigger than $k_{\rm max}(i)$ which we interpret as a signal of a Big Crunch. In fact, this expectations will be supported by the findings from the next subsection.

\subsection{Evolution of gaussian physical states}\label{sc:evolution_of_gaussian_states}
\begin{figure}[!tbp]
\centering
\includegraphics{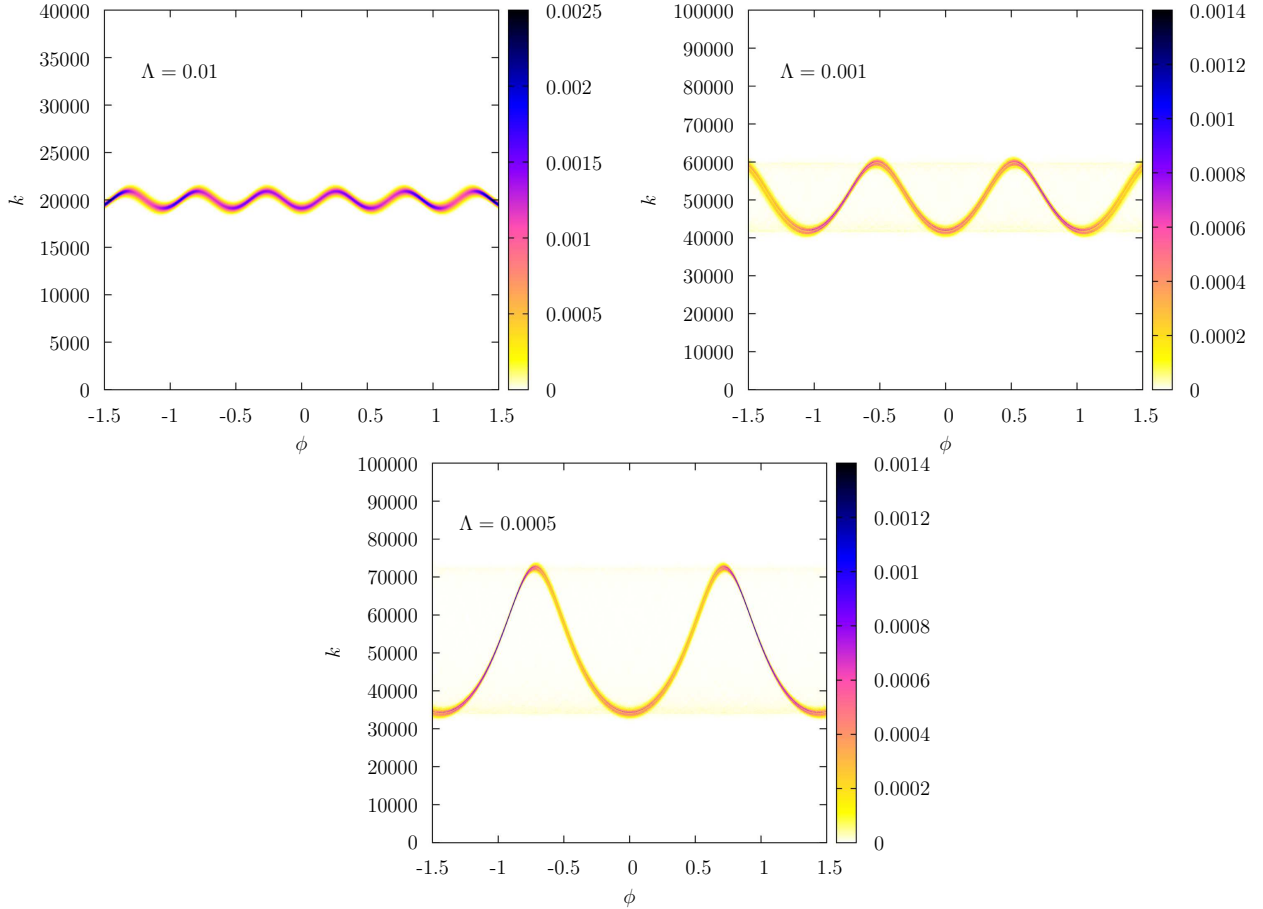}
\caption{We plotted the probability density $|\Psi(k,\phi)|^2$ of the gaussian coherent state \eqref{eq:gaussian_coherent_state} for three values of cosmological constant $\Lambda=0.01,0.001,0.0005$.}\label{fig:coherent_states}
\end{figure}
Physical states are of the form:
\be\label{eq:evolution}
\Psi(k,\phi)=\sum_{i: \Theta_i < 0}  \widetilde{\Psi}_i(k) e^{\iu\, \omega_i \phi}  e_{i}(k),
\ee
where $\tilde{\Psi}(k)$ is a profile function, 
\be 
\omega_i=\sqrt{-\Theta_i}
\ee
and the sum is over $i$ such that $\Theta_i< 0$. Let us recall that $\Theta_i$ is the $i$-th eigenvalue of the operator $\hat{\Theta}_{\Lambda}$ (ordered in an increasing order).

In the original case of open Universe with zero cosmological constant, the sum in the formula \eqref{eq:evolution} is replaced by an integral (see for example \cite{LQCIV,LQCII,LQCIII}). However, in the case of a closed Universe studied in \cite{AshtekarPawlowskiClosedUniverse} the spectrum was discrete and the integral was replaced by a sum. We expect that also in our case the eigenstates are normalizable and the spectrum is discrete. We base our expectations on a numerical study of the spectral properties of our quantum Hamiltonian which was done in section \ref{sc:spectral_properties}.
%Let us notice that $\omega_i \phi$ is dimensionless. Therefore the formula \eqref{eq:evolution} holds in any units. 

We calculated the evolution of the gaussian physical state:
\be\label{eq:gaussian_coherent_state}
\Psi(k,\phi)=\sum_{i} \frac{1}{N_{\rm G}} e^{-\frac{(i-i_0)}{2 \sigma^2}} e^{\iu\, \omega_i \phi} e_{i}(k)  ,
\ee
where $N_{\rm G}$ is the normalization factor. In practical numerical calculations we limited the range of the sum to $i$ for which 
\be
|i-i_0 | \leq 5 \sigma.
\ee
Let us notice that for $i_{\rm ext}$ such that $|i_{\rm ext}-i_0 |=5\sigma$ the factor $e^{-\frac{(i-i_0)}{2 \sigma^2}}$ is $e^{-25/2}\approx 3.727\cdot 10^{-9}$ which is below the machine epsilon for single precision arithmetics. In our calculations we chose the rank of the truncated matrix to be big enough to ensure that for each $i$ in the range we considered the eigenvector $e_i$ and eigenvalue $\Theta_i$ converged. On figure \ref{fig:coherent_states} we presented the plots of the probability density $|\Psi(k,\phi)|^2$ in three cases $\Lambda=0.01, 0.001, 0.0005$:
\begin{enumerate}
\item In the case $\Lambda=0.01$ the matrix rank was $40000$, the coherent state \eqref{eq:gaussian_coherent_state} was peaked at $i_0=20000$ and the standard deviation was $\sigma=400$. This corresponds to 
\be
p_{\phi}\approx 1.6 \cdot 10^{5}, \frac{\Delta p_{\phi}}{p_{\phi}}\approx0.03.
\ee
\item In the case $\Lambda=0.001$ the matrix rank was $100000$, the coherent state \eqref{eq:gaussian_coherent_state} was peaked at $i_0=50000$ and the standard deviation was $\sigma=1000$. This corresponds to 
\be 
p_{\phi}\approx 1.9 \cdot 10^{5}, \frac{\Delta p_{\phi}}{p_{\phi}}\approx0.03.
\ee

\item In the case $\Lambda=0.0005$, the matrix rank was $100000$ the coherent state \eqref{eq:gaussian_coherent_state} was peaked at $i_0=50000$ and the standard deviation was $\sigma=1000$. This corresponds to 
\be 
p_{\phi}\approx 1.25 \cdot 10^{5}, \frac{\Delta p_{\phi}}{p_{\phi}}\approx0.03.
\ee
\end{enumerate} 
\begin{figure}[!tbp]
\centering
\includegraphics{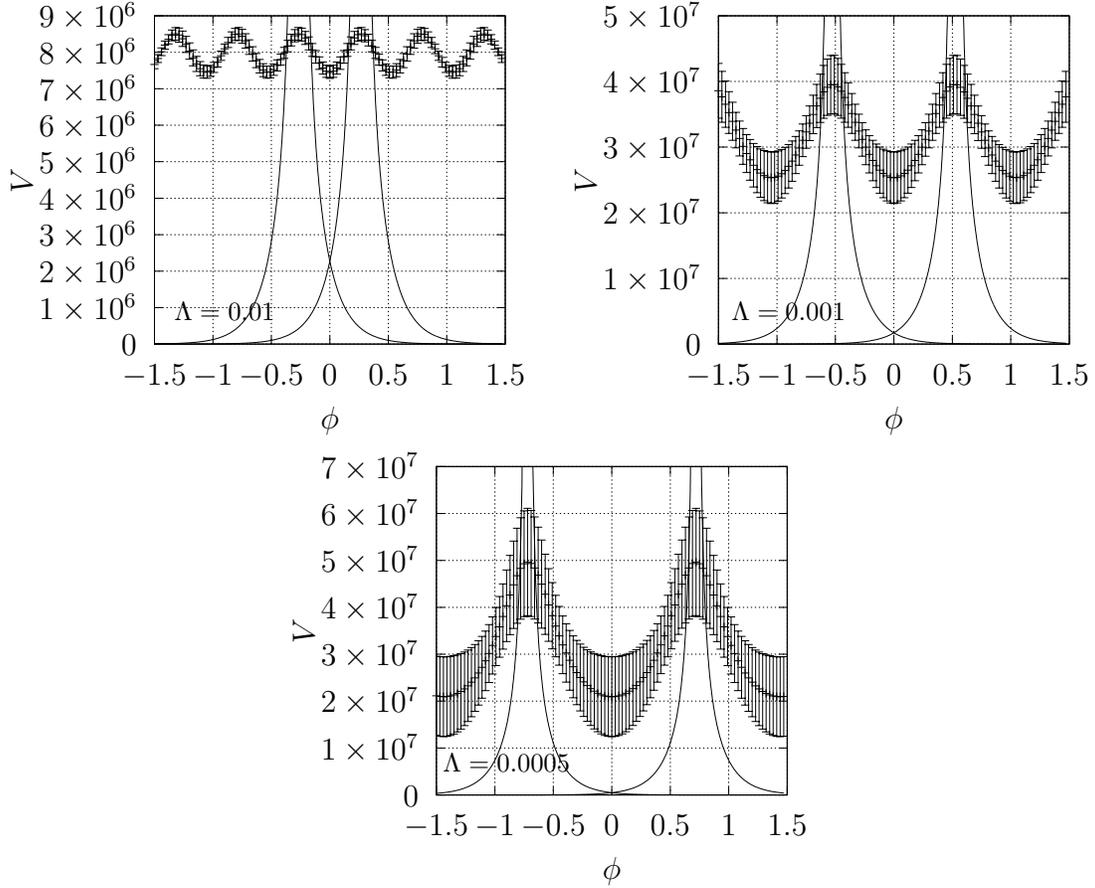}
\caption{We plotted the volume expectation values and their dispersion for the gaussian coherent states \eqref{eq:gaussian_coherent_state} as a function of $\phi$. We considered three values of cosmological constant $\Lambda=0.01,0.001,0.0005$. The solid lines represent two corresponding classical solutions.}\label{fig:volume_expectation_values}
\end{figure}

The classical evolution of the system is given by \cite{AshtekarPawlowskiCosmologicalConstant}:
\be\label{eq:classical_evolution}
V(\phi)=\frac{\sqrt{4\pi} p_\phi}{\sqrt{\Lambda}} \frac{1}{|\sinh(\sqrt{12\pi(\phi-\phi_0)})|}.
\ee
We calculated the volume expectation values and their dispersions for different values of $\phi$. The result of our calculation is plotted on figure \ref{fig:volume_expectation_values}. With a solid line we denoted two classical curves \eqref{eq:classical_evolution} corresponding to two values of $\phi_0$ located at two local maxima nearby the local minimum at $\phi=0$.
\section{Summary}
We selected the zero internal curvature sector of a homogeneous-isotropic space proposed in \cite{HomogeneousIsotropic}. The space can be constructed in the following way. In \cite{HomogeneousIsotropic} we introduced a space $\Hil^{\rm loops}_{\Gamma,j,x}$ which is spanned by vectors of the form:
\begin{equation}
\ket{\loopconfiguration,\iota},
\end{equation}
where $\loopconfiguration$ runs through all possible loop configurations (see section \ref{sc:coherent_states}, in particular figure \ref{fig:loop_configuration_ab}) and $\iota$ runs through a basis of certain intertwiners. In our approach each link of the corresponding spin-network is labeled with spin $\frac{1}{2}$ (i.e. each side of the lattice and each loop). Let us underline, that the states $\ket{\loopconfiguration,\iota}$ corresponding to different loop configurations are orthogonal
\begin{equation}
\braket{\loopconfiguration',\iota'}{\loopconfiguration,\iota}=\delta_{\loopconfiguration',\loopconfiguration} \braket{\iota'}{\iota}.
\end{equation}
For each $\loopconfiguration$ we construct a unique intertwiner $\iota_\loopconfiguration$ which is built from certain Livine-Speziale intertwiners. This leads us to a subspace spanned by vectors
\begin{equation}
\ket{\loopconfiguration}=\ket{\loopconfiguration,\iota_\loopconfiguration}.
\end{equation}
The states do not have the necessary symmetry properties. Following \cite{HomogeneousIsotropic} we project them onto a subspace of homogeneous-isotropic states by averaging over a group of discrete rotations $\CubicalSymmetries$ (orientation preserving symmetries of a cube):
\begin{equation}
\ket{[\loopconfiguration]}=\frac{1}{\sqrt{\# \orbit{\loopconfiguration}}} \sum_{\loopconfiguration'\in \orbit{\loopconfiguration}} \ket{\loopconfiguration'},
\end{equation}
where $\orbit{\loopconfiguration}$ denotes the orbit of the loop configuration $\loopconfiguration$. Finally, for each $k\in \mathbb{N}$ we specify in a unique way a loop configuration $\loopconfiguration_k$ with $k$ loops (see section \ref{sc:basis_states}), i.e. $\numberofloops(\loopconfiguration_k)=k$. The choice is made in a way which makes the calculations easier. Finally, our cosmological Hilbert space, which we denoted by $\Hil_{\rm flat}$, is spanned by the following states:
\begin{equation}
\ket{k}:=\ket{[\loopconfiguration_k]}.
\end{equation}
In section \ref{sc:internal_flatness} we argued that the states encode quantum geometries with zero intrinsic curvature (which is shown by studying the expectation values of the dihedral angle operator). The resulting Hilbert space resembles the Loop Quantum Cosmology Hilbert space. In fact, thanks to the result from \cite{VolumeMonochromatic} the states are eigenstates of the volume operator:
\begin{equation}
\hat{V}\ \ket{k}=\frac{\kappa_0}{8} \left(\frac{8\pi G \hbar \gamma}{c^3}\right)^{\frac{3}{2}} \sqrt{\frac{4\sqrt{3}}{3} (k+2)(k+3)(k+4)}\ \ket{k}.
\end{equation}
It is clear that the correspondence with the standard formulation of LQC is not complete, because the basis is not equally spaced in the volume eigenvalues but rather in the number of loops $k$. 

After constructing the states, we project the quantum Hamiltonian of Loop Quantum Gravity to our space. Strictly speaking, we consider an operator $\hat{\Theta}_{\Lambda}$ defined on (a dense domain of) $\Hil_{\rm flat}$. It has the following defining property. For any pair of states $\ket{\Psi}$ and $\ket{\Psi'}$ which are finite linear combinations of the states $\ket{k}$ the matrix elements of the operator $\hat{\Theta}_{\Lambda}$ coincide with the matrix elements of the gravitational part of scalar constraint operator $\widehat{\sqrt{q} C^{\rm gr}}$:
\begin{equation}\label{eq:summary_correspondence}
\bra{\Psi'}\hat{\Theta}_{\Lambda}\ket{\Psi}\approx \bra{\Psi'}\widehat{\sqrt{q} C^{\rm gr}}\ket{\Psi}.
\end{equation}
Our Hamiltonian is 
\be 
\hat{H}=\sqrt{-\hat{\Theta}_\Lambda}.
\ee
The correspondence \eqref{eq:summary_correspondence} looks very similar to the requirement for a relation between Loop Quantum Cosmology and Loop Quantum Gravity proposed in \cite{EngleLQCI,EngleLQCII,EngleLQCIII}. However, in our model the relation cannot be considered to be complete, because our reduced model is defined on a different Hilbert space than the standard LQC models. We intentionally used the symbol $\approx$ in \eqref{eq:summary_correspondence} to underline that the equality holds only approximately. We made some simplifying assumptions, which are discussed in section \ref{sc:euclidean_part}. Let us recall what simplifications we made. We omitted the so-called Lorentzian part of the gravitational scalar constraint operator corresponding to operator $\widehat{\int d^3x \sqrt{q}^2 R^{(3)}}$ (because the quantum geometries are intrinsically flat -- see section \ref{sc:internal_flatness} in particular figure \ref{fig:dihedral_angle}) and we approximate the matrix elements of the Euclidean part using a formula based on the stationary phase analysis improved by our numerical experiments (see sections \ref{sc:euclidean_asymptotic} and \ref{sc:euclidean_numerical}, in particular figure \ref{fig:CE_matrix_elements}). 

In section \ref{sc:semiclassical} we argued that our operator $\hat{\Theta}_\Lambda$ has a proper semi-classical limit (in the region, where $k$ is large but not too large). We investigated the spectral properties of the operator numerically in section \ref{sc:spectral_properties}. We noticed that each eigenvector is localized in a region of $k$. Basing on arguments from \cite{ZhangBouncingUniverse} we could expect that the model will experience a Big Bounce and a Big Crunch. We confirmed this by studying evolution of Gaussian states (constructed in complete analogy to LQC) -- see section \ref{sc:evolution_of_gaussian_states}, in particular figures \ref{fig:coherent_states} and \ref{fig:volume_expectation_values}. As a result, the Universe in our model experiences a periodic evolution: Big Bounce is followed by a classical expansion region, which is followed by Big Crunch, which is followed by classical contracting region, which is followed by a Big Bounce etc. This is in accord with the results in Loop Quantum Cosmology \cite{AshtekarPawlowskiCosmologicalConstant}.

\section{Discussion and outlook}
The results from this paper have important conceptual consequences. In our model the symmetry reduction is done at the quantum level. As a result, any test of the reduced theory would be a test of the full Loop Quantum Gravity theory. Let us notice that any physical prediction of Loop Quantum Cosmology (such as deviations in the classical CMB power spectrum \cite{AshtekarCMBI,AshtekarCMBII,AshtekarCMBIII}) supports or denies LQC but has only indirect consequences for Loop Quantum Gravity. In addition, Loop Quantum Cosmology applies to cosmological models and has limited application to other symmetry reduction schemes (used for example in the study of black-hole physics). We expect that the techniques developed in our research can be applied to other symmetry reduced models, possibly leading to further tests of the same full theory.

The results have also important technical consequences. Although, we restrict to homogeneous-isotropic sector, inhomogeneities and anisotropies can be taken into account. For example considering small perturbations of the homogeneous-isotropic geometries could be done now in a single scheme. Let us recall in the recent proposal in Loop Quantum Cosmology \cite{AshtekarCMBI,AshtekarCMBII,AshtekarCMBIII} the background is quantized using Loop Quantum Cosmology quantization but perturbations are quantized using the Fock quantization. Moreover, our approach introduces a natural splitting of the quantum Hamiltonian into homogeneous-isotropic part and the rest, which suggests that in our approach the technical tool appropriate for this problem is the perturbation theory of operators. Recently there has been some development in this direction in the context of Loop Quantum Cosmology \cite{AshtekarSpinFoamI,AshtekarSpinFoamII,AshtekarSpinFoamIII} and Loop Quantum Gravity \cite{SFscalar}.

The cosmological constant plays a very important role in our approach. The resulting model experiences Big Bounce and Big Crunch. Both (hypothetical) phenomena are due to Quantum Gravity effects and the classical regime is in between. The smaller cosmological constant, the larger is the classical region and the larger is the maximal size of the Universe. As a result, in order to be compatible with the current observations, i.e.  classical Universe of very large size in Planck units, we need a very small (but positive) cosmological constant in our model. The particular numerical values need a further study (see technical remarks at the end of this section). 

We shed some new light on the relation between LQC and LQG. In this paper we projected the full Loop Quantum Gravity Hamiltonian onto our space of homogeneous-isotropic states corresponding to zero intrinsic curvature geometries. The resulting operator resembles the Loop Quantum Cosmology Hamiltonian. The advantage of our approach is that the states are naturally embedded in the full LQG Hilbert space and the matrix elements of the full Loop Quantum Gravity Hamiltonian coincide with matrix elements of the quantum Hamiltonian in the reduced theory. These are the conditions proposed in \cite{EngleLQCI,EngleLQCII,EngleLQCIII} as a proper correspondence between LQC and LQG. However, in our approach, the reduced quantum theory is not the LQC in the standard formulation. The crucial difference between the LQC Hilbert space and our Hilbert space is that in LQC it is build from equally spaced volume eigenstates. In our approach we use a basis equally spaced in the number of loops. Our states are also eigenstates of the volume operator but the relation between the number of loops and volume is non-linear (see \eqref{eq:volume_monochromatic}). For completeness of the correspondence, one should look for an alternative LQC quantization realized on our space.

There are a number of technical problems which need a further study. We will name some of them:
\begin{itemize}
\item Our zero-intrinsic curvature space $\Hil_{\rm flat}$ is not preserved by the quantum Hamiltonian of the full theory. It would be interesting to verify if the space is preserved in some approximate sense. For example, it would be interesting to check if the eigenvectors from the reduced theory solve the full eigenvalue equation approximately.
\item For each $k\in \mathbb{N}$ we chose a loop configuration $\loopconfiguration_k$ with $k$ loops (see  section \ref{sc:basis_states}). This choice was made to simplify the calculations. An alternative would be to consider states $\ket{k}$ as some other combination of states $\ket{\loopconfiguration}$ such that $\numberofloops(\loopconfiguration)=k$. We believe that one proposal is particularly worth investigating. Let us describe it shortly. Let us notice that there are $15$ pairs of sides of a cubical lattice meeting at a given node $n$ but $3$ pairs are antiparallel. This gives $12$ pairs $(I,J), I<J$ that appear in the domain of a loop configuration. We can order them (lexicographically) and number with indices in the set $\{1,\ldots,12\}$. Let us denote by $\alpha$ a function that assigns to a pair of links its corresponding index in $\{1,\ldots,12\}$. Let us consider the following function:
\begin{equation}
f_k(x_1,x_2,\ldots,x_{12})=(x_1+\ldots+x_{12})^k=\sum_{\substack{n_1,\ldots, n_{12}:\\ n_1+\ldots+n_{12}=k } } \binom{k}{n_1,\ldots,n_{12}}\cdot x_1^{n_1}\cdot x_2^{n_2}\cdot \ldots \cdot x_{12}^{n_{12}}.
\end{equation}
Clearly, in the formula above:
\begin{equation}
\binom{k}{n_1,\ldots,n_{12}}=\frac{k!}{ n_1! n_2! \ldots n_{12}!}
\end{equation}
is the multinomial coefficient. We construct an alternative state to our state $\ket{k}$ in the following way. Firstly, we construct a state 
\be 
\widetilde{\ket{k}}=\sum_{\substack{n_1,\ldots, n_{12}:\\ n_1+\ldots+n_{12}=k } } \binom{k}{n_1,\ldots,n_{12}} \ket{\loopconfiguration_{n_1,\ldots,n_{12}}},
\ee
where $\loopconfiguration_{n_1,\ldots,n_{12}}$ is a loop configuration such that $\loopconfiguration_{n_1,\ldots,n_{12}}(I,J)=n_{\alpha(I,J)}$. The alternative state to $\ket{k}$ is obtained by normalizing the state $\widetilde{\ket{k}}$. The resulting state has interesting properties. First of all, it is homogeneous-isotropic. The action of element $g\in\CubicalSymmetries$ induces an action of $g$ on the pairs of links which results in a permutation of the numbers $\{1,\ldots,12\}$. Since the function $f_k$ is permutation invariant, the resulting state is homogeneous-isotropic. Moreover, 
\be
f_{k+1}(x_1,x_2,\ldots,x_{12})=f_{k}(x_1,x_2,\ldots,x_{12})\cdot (x_1+\ldots+x_{12}).
\ee 
This has an effect that each $\CE^{\dagger}_{IJ}$ contributes with the same factor to the matrix elements of $\CE^{\dagger}$ in such states. Let us recall that in the proposal studied in this paper only certain $\CE^{\dagger}_{IJ}$ contributed and we needed to consider the three cases $k=3n, k=3n+1, k=3n+2$ separately.

We expect that the resulting cosmological Hamiltonian will not differ much but the Hilbert space $\Hil_{\rm flat}$ will be different and its properties may change. This is important for example in the study of the previous technical problem, i.e. the study of approximate invariance of $\Hil_{\rm flat}$. Such choices probably lead to more complicated calculations and we leave them for future research. 
\item Our results need more detailed numerical study. Due to limitations of the numerical methods that we used, we were able to study only moderately small cosmological constants: $\Lambda=0.0005$ was the smallest. As a result, the classical region was barely visible (on figure \ref{fig:volume_expectation_values} we compare the results of the quantum evolution with the classical trajectory). Smaller values of cosmological constant need to be considered but this involves larger matrices and is computationally more demanding. 
\item We fixed the free constant $\kappa_0$ in the volume operator to be $1$. As far as we know, the value of this free constant can be fixed in the case of Ashtekar-Lewandowski volume operator \cite{AshtekarLewandowskiVolume, ThiemannConsisencyCheckI,ThiemannConsisencyCheckII} but not the Rovelli-Smollin volume operator. Since the cosmological constant enters the formulas together with the factor $\kappa_0^2$, it is important to fix this parameter in order to get a proper scale of the cosmological constant. This problem may re-appear when some fine-graining procedure will be studied. In our model, we implicitly assumed that a fundamental cell contains only one node. When a fundamental cell contains $N$ nodes the volume gets multiplied by $N^3$. This factor could be also incorporated in $\kappa_0$.
\end{itemize}

\section*{Acknowledgements}
I would like to thank Simone Speziale and Alejandro Perez from Centre de Physique Theorique, Aix-Marseille University for warm hospitality during my research visit and for stimulating discussions. This work was supported by the National Science Centre, Poland grant No. 2018/28/C/ST9/00157. 
\appendix
\section*{References}
\bibliography{LQG}{}

%merlin.mbs apsrev4-1.bst 2010-07-25 4.21a (PWD, AO, DPC) hacked
%Control: key (0)
%Control: author (8) initials jnrlst
%Control: editor formatted (1) identically to author
%Control: production of article title (-1) disabled
%Control: page (0) single
%Control: year (1) truncated
%Control: production of eprint (0) enabled
\begin{thebibliography}{54}%
\makeatletter
\providecommand \@ifxundefined [1]{%
 \@ifx{#1\undefined}
}%
\providecommand \@ifnum [1]{%
 \ifnum #1\expandafter \@firstoftwo
 \else \expandafter \@secondoftwo
 \fi
}%
\providecommand \@ifx [1]{%
 \ifx #1\expandafter \@firstoftwo
 \else \expandafter \@secondoftwo
 \fi
}%
\providecommand \natexlab [1]{#1}%
\providecommand \enquote  [1]{``#1''}%
\providecommand \bibnamefont  [1]{#1}%
\providecommand \bibfnamefont [1]{#1}%
\providecommand \citenamefont [1]{#1}%
\providecommand \href@noop [0]{\@secondoftwo}%
\providecommand \href [0]{\begingroup \@sanitize@url \@href}%
\providecommand \@href[1]{\@@startlink{#1}\@@href}%
\providecommand \@@href[1]{\endgroup#1\@@endlink}%
\providecommand \@sanitize@url [0]{\catcode `\\12\catcode `\$12\catcode
  `\&12\catcode `\#12\catcode `\^12\catcode `\_12\catcode `\%12\relax}%
\providecommand \@@startlink[1]{}%
\providecommand \@@endlink[0]{}%
\providecommand \url  [0]{\begingroup\@sanitize@url \@url }%
\providecommand \@url [1]{\endgroup\@href {#1}{\urlprefix }}%
\providecommand \urlprefix  [0]{URL }%
\providecommand \Eprint [0]{\href }%
\providecommand \doibase [0]{http://dx.doi.org/}%
\providecommand \selectlanguage [0]{\@gobble}%
\providecommand \bibinfo  [0]{\@secondoftwo}%
\providecommand \bibfield  [0]{\@secondoftwo}%
\providecommand \translation [1]{[#1]}%
\providecommand \BibitemOpen [0]{}%
\providecommand \bibitemStop [0]{}%
\providecommand \bibitemNoStop [0]{.\EOS\space}%
\providecommand \EOS [0]{\spacefactor3000\relax}%
\providecommand \BibitemShut  [1]{\csname bibitem#1\endcsname}%
\let\auto@bib@innerbib\@empty
%</preamble>
\bibitem [{\citenamefont {Ashtekar}\ and\ \citenamefont
  {Lewandowski}(2004)}]{StatusReport}%
  \BibitemOpen
  \bibfield  {author} {\bibinfo {author} {\bibfnamefont {A.}~\bibnamefont
  {Ashtekar}}\ and\ \bibinfo {author} {\bibfnamefont {J.}~\bibnamefont
  {Lewandowski}},\ }\href {\doibase 10.1088/0264-9381/21/15/R01} {\bibfield
  {journal} {\bibinfo  {journal} {Class.Quant.Grav.}\ }\textbf {\bibinfo
  {volume} {21}},\ \bibinfo {pages} {R53} (\bibinfo {year} {2004})},\ \Eprint
  {http://arxiv.org/abs/gr-qc/0404018} {arXiv:gr-qc/0404018 [gr-qc]}
  \BibitemShut {NoStop}%
%%CITATION = GR-QC/0404018;%%
\bibitem [{\citenamefont {Thiemann}(2007)}]{ThiemannBook}%
  \BibitemOpen
  \bibfield  {author} {\bibinfo {author} {\bibfnamefont {T.}~\bibnamefont
  {Thiemann}},\ }\href@noop {} {\emph {\bibinfo {title} {{Modern Canonical
  Quantum General Relativity}}}}\ (\bibinfo  {publisher} {Cambridge University
  Press},\ \bibinfo {year} {2007})\BibitemShut {NoStop}%
\bibitem [{\citenamefont {Rovelli}(2004)}]{RovelliBook}%
  \BibitemOpen
  \bibfield  {author} {\bibinfo {author} {\bibfnamefont {C.}~\bibnamefont
  {Rovelli}},\ }\href@noop {} {\emph {\bibinfo {title} {{Quantum Gravity}}}}\
  (\bibinfo  {publisher} {Cambridge University Press},\ \bibinfo {year}
  {2004})\BibitemShut {NoStop}%
\bibitem [{\citenamefont {Rovelli}(2011{\natexlab{a}})}]{Zakopane}%
  \BibitemOpen
  \bibfield  {author} {\bibinfo {author} {\bibfnamefont {C.}~\bibnamefont
  {Rovelli}},\ }\href@noop {} {\bibfield  {journal} {\bibinfo  {journal} {PoS}\
  }\textbf {\bibinfo {volume} {QGQGS2011}},\ \bibinfo {pages} {003} (\bibinfo
  {year} {2011}{\natexlab{a}})},\ \Eprint {http://arxiv.org/abs/1102.3660}
  {arXiv:1102.3660 [gr-qc]} \BibitemShut {NoStop}%
%%CITATION = ARXIV:1102.3660;%%
\bibitem [{\citenamefont {Rovelli}(2011{\natexlab{b}})}]{LQG25}%
  \BibitemOpen
  \bibfield  {author} {\bibinfo {author} {\bibfnamefont {C.}~\bibnamefont
  {Rovelli}},\ }\href {\doibase 10.1088/0264-9381/28/15/153002} {\bibfield
  {journal} {\bibinfo  {journal} {Class.Quant.Grav.}\ }\textbf {\bibinfo
  {volume} {28}},\ \bibinfo {pages} {153002} (\bibinfo {year}
  {2011}{\natexlab{b}})},\ \Eprint {http://arxiv.org/abs/1012.4707}
  {arXiv:1012.4707 [gr-qc]} \BibitemShut {NoStop}%
%%CITATION = ARXIV:1012.4707;%%
\bibitem [{\citenamefont {Han}\ \emph {et~al.}(2007)\citenamefont {Han},
  \citenamefont {Huang},\ and\ \citenamefont {Ma}}]{MaReview}%
  \BibitemOpen
  \bibfield  {author} {\bibinfo {author} {\bibfnamefont {M.}~\bibnamefont
  {Han}}, \bibinfo {author} {\bibfnamefont {W.}~\bibnamefont {Huang}}, \ and\
  \bibinfo {author} {\bibfnamefont {Y.}~\bibnamefont {Ma}},\ }\href {\doibase
  10.1142/S0218271807010894} {\bibfield  {journal} {\bibinfo  {journal}
  {Int.J.Mod.Phys.}\ }\textbf {\bibinfo {volume} {D16}},\ \bibinfo {pages}
  {1397} (\bibinfo {year} {2007})},\ \Eprint
  {http://arxiv.org/abs/gr-qc/0509064} {arXiv:gr-qc/0509064 [gr-qc]}
  \BibitemShut {NoStop}%
%%CITATION = GR-QC/0509064;%%
\bibitem [{\citenamefont {Ashtekar}\ \emph {et~al.}(2014)\citenamefont
  {Ashtekar}, \citenamefont {Reuter},\ and\ \citenamefont
  {Rovelli}}]{AshtekarRovelliReuterRev}%
  \BibitemOpen
  \bibfield  {author} {\bibinfo {author} {\bibfnamefont {A.}~\bibnamefont
  {Ashtekar}}, \bibinfo {author} {\bibfnamefont {M.}~\bibnamefont {Reuter}}, \
  and\ \bibinfo {author} {\bibfnamefont {C.}~\bibnamefont {Rovelli}},\
  }\href@noop {} {\  (\bibinfo {year} {2014})},\ \Eprint
  {http://arxiv.org/abs/1408.4336} {arXiv:1408.4336 [gr-qc]} \BibitemShut
  {NoStop}%
%%CITATION = ARXIV:1408.4336;%%
\bibitem [{\citenamefont {Giesel}\ and\ \citenamefont
  {Thiemann}(2010)}]{AQGIV}%
  \BibitemOpen
  \bibfield  {author} {\bibinfo {author} {\bibfnamefont {K.}~\bibnamefont
  {Giesel}}\ and\ \bibinfo {author} {\bibfnamefont {T.}~\bibnamefont
  {Thiemann}},\ }\href {\doibase 10.1088/0264-9381/27/17/175009} {\bibfield
  {journal} {\bibinfo  {journal} {Class. Quant. Grav.}\ }\textbf {\bibinfo
  {volume} {27}},\ \bibinfo {pages} {175009} (\bibinfo {year} {2010})},\
  \Eprint {http://arxiv.org/abs/0711.0119} {arXiv:0711.0119 [gr-qc]}
  \BibitemShut {NoStop}%
%%CITATION = ARXIV:0711.0119;%%
\bibitem [{\citenamefont {Domagala}\ \emph {et~al.}(2010)\citenamefont
  {Domagala}, \citenamefont {Giesel}, \citenamefont {Kaminski},\ and\
  \citenamefont {Lewandowski}}]{GravityQuantized}%
  \BibitemOpen
  \bibfield  {author} {\bibinfo {author} {\bibfnamefont {M.}~\bibnamefont
  {Domagala}}, \bibinfo {author} {\bibfnamefont {K.}~\bibnamefont {Giesel}},
  \bibinfo {author} {\bibfnamefont {W.}~\bibnamefont {Kaminski}}, \ and\
  \bibinfo {author} {\bibfnamefont {J.}~\bibnamefont {Lewandowski}},\ }\href
  {\doibase 10.1103/PhysRevD.82.104038} {\bibfield  {journal} {\bibinfo
  {journal} {Phys. Rev.}\ }\textbf {\bibinfo {volume} {D82}},\ \bibinfo {pages}
  {104038} (\bibinfo {year} {2010})},\ \Eprint {http://arxiv.org/abs/1009.2445}
  {arXiv:1009.2445 [gr-qc]} \BibitemShut {NoStop}%
%%CITATION = ARXIV:1009.2445;%%
\bibitem [{\citenamefont {Bojowald}(2008)}]{LQCI}%
  \BibitemOpen
  \bibfield  {author} {\bibinfo {author} {\bibfnamefont {M.}~\bibnamefont
  {Bojowald}},\ }\href@noop {} {\bibfield  {journal} {\bibinfo  {journal}
  {Living Rev. Rel.}\ }\textbf {\bibinfo {volume} {11}},\ \bibinfo {pages} {4}
  (\bibinfo {year} {2008})}\BibitemShut {NoStop}%
%%CITATION = 00222,11,4;%%
\bibitem [{\citenamefont {Ashtekar}\ and\ \citenamefont {Singh}(2011)}]{LQCII}%
  \BibitemOpen
  \bibfield  {author} {\bibinfo {author} {\bibfnamefont {A.}~\bibnamefont
  {Ashtekar}}\ and\ \bibinfo {author} {\bibfnamefont {P.}~\bibnamefont
  {Singh}},\ }\href {\doibase 10.1088/0264-9381/28/21/213001} {\bibfield
  {journal} {\bibinfo  {journal} {Class. Quant. Grav.}\ }\textbf {\bibinfo
  {volume} {28}},\ \bibinfo {pages} {213001} (\bibinfo {year} {2011})},\
  \Eprint {http://arxiv.org/abs/1108.0893} {arXiv:1108.0893 [gr-qc]}
  \BibitemShut {NoStop}%
%%CITATION = ARXIV:1108.0893;%%
\bibitem [{\citenamefont {Agullo}\ and\ \citenamefont {Singh}(2017)}]{LQCIII}%
  \BibitemOpen
  \bibfield  {author} {\bibinfo {author} {\bibfnamefont {I.}~\bibnamefont
  {Agullo}}\ and\ \bibinfo {author} {\bibfnamefont {P.}~\bibnamefont {Singh}},\
  }in\ \href {\doibase 10.1142/9789813220003_0007} {\emph {\bibinfo {booktitle}
  {Loop Quantum Gravity: The First 30 Years}}},\ \bibinfo {editor} {edited by\
  \bibinfo {editor} {\bibfnamefont {A.}~\bibnamefont {Ashtekar}}\ and\ \bibinfo
  {editor} {\bibfnamefont {J.}~\bibnamefont {Pullin}}}\ (\bibinfo  {publisher}
  {WSP},\ \bibinfo {year} {2017})\ pp.\ \bibinfo {pages} {183--240},\ \Eprint
  {http://arxiv.org/abs/1612.01236} {arXiv:1612.01236 [gr-qc]} \BibitemShut
  {NoStop}%
%%CITATION = ARXIV:1612.01236;%%
\bibitem [{\citenamefont {Ashtekar}\ \emph {et~al.}(2006)\citenamefont
  {Ashtekar}, \citenamefont {Pawlowski},\ and\ \citenamefont {Singh}}]{LQCIV}%
  \BibitemOpen
  \bibfield  {author} {\bibinfo {author} {\bibfnamefont {A.}~\bibnamefont
  {Ashtekar}}, \bibinfo {author} {\bibfnamefont {T.}~\bibnamefont {Pawlowski}},
  \ and\ \bibinfo {author} {\bibfnamefont {P.}~\bibnamefont {Singh}},\ }\href
  {\doibase 10.1103/PhysRevLett.96.141301} {\bibfield  {journal} {\bibinfo
  {journal} {Phys. Rev. Lett.}\ }\textbf {\bibinfo {volume} {96}},\ \bibinfo
  {pages} {141301} (\bibinfo {year} {2006})},\ \Eprint
  {http://arxiv.org/abs/gr-qc/0602086} {arXiv:gr-qc/0602086 [gr-qc]}
  \BibitemShut {NoStop}%
%%CITATION = GR-QC/0602086;%%
\bibitem [{\citenamefont {Alesci}\ and\ \citenamefont
  {Cianfrani}(2013{\natexlab{a}})}]{QRLQGI}%
  \BibitemOpen
  \bibfield  {author} {\bibinfo {author} {\bibfnamefont {E.}~\bibnamefont
  {Alesci}}\ and\ \bibinfo {author} {\bibfnamefont {F.}~\bibnamefont
  {Cianfrani}},\ }\href {\doibase 10.1209/0295-5075/104/10001} {\bibfield
  {journal} {\bibinfo  {journal} {EPL}\ }\textbf {\bibinfo {volume} {104}},\
  \bibinfo {pages} {10001} (\bibinfo {year} {2013}{\natexlab{a}})},\ \Eprint
  {http://arxiv.org/abs/1210.4504} {arXiv:1210.4504 [gr-qc]} \BibitemShut
  {NoStop}%
%%CITATION = ARXIV:1210.4504;%%
\bibitem [{\citenamefont {Alesci}\ and\ \citenamefont
  {Cianfrani}(2013{\natexlab{b}})}]{QRLQGII}%
  \BibitemOpen
  \bibfield  {author} {\bibinfo {author} {\bibfnamefont {E.}~\bibnamefont
  {Alesci}}\ and\ \bibinfo {author} {\bibfnamefont {F.}~\bibnamefont
  {Cianfrani}},\ }\href {\doibase 10.1103/PhysRevD.87.083521} {\bibfield
  {journal} {\bibinfo  {journal} {Phys. Rev.}\ }\textbf {\bibinfo {volume}
  {D87}},\ \bibinfo {pages} {083521} (\bibinfo {year} {2013}{\natexlab{b}})},\
  \Eprint {http://arxiv.org/abs/1301.2245} {arXiv:1301.2245 [gr-qc]}
  \BibitemShut {NoStop}%
%%CITATION = ARXIV:1301.2245;%%
\bibitem [{\citenamefont {Alesci}\ \emph {et~al.}(2013)\citenamefont {Alesci},
  \citenamefont {Cianfrani},\ and\ \citenamefont {Rovelli}}]{QRLQGIII}%
  \BibitemOpen
  \bibfield  {author} {\bibinfo {author} {\bibfnamefont {E.}~\bibnamefont
  {Alesci}}, \bibinfo {author} {\bibfnamefont {F.}~\bibnamefont {Cianfrani}}, \
  and\ \bibinfo {author} {\bibfnamefont {C.}~\bibnamefont {Rovelli}},\ }\href
  {\doibase 10.1103/PhysRevD.88.104001} {\bibfield  {journal} {\bibinfo
  {journal} {Phys. Rev.}\ }\textbf {\bibinfo {volume} {D88}},\ \bibinfo {pages}
  {104001} (\bibinfo {year} {2013})},\ \Eprint {http://arxiv.org/abs/1309.6304}
  {arXiv:1309.6304 [gr-qc]} \BibitemShut {NoStop}%
%%CITATION = ARXIV:1309.6304;%%
\bibitem [{\citenamefont {Alesci}\ and\ \citenamefont
  {Cianfrani}(2014)}]{QRLQGIV}%
  \BibitemOpen
  \bibfield  {author} {\bibinfo {author} {\bibfnamefont {E.}~\bibnamefont
  {Alesci}}\ and\ \bibinfo {author} {\bibfnamefont {F.}~\bibnamefont
  {Cianfrani}},\ }\href {\doibase 10.1103/PhysRevD.90.024006} {\bibfield
  {journal} {\bibinfo  {journal} {Phys. Rev.}\ }\textbf {\bibinfo {volume}
  {D90}},\ \bibinfo {pages} {024006} (\bibinfo {year} {2014})},\ \Eprint
  {http://arxiv.org/abs/1402.3155} {arXiv:1402.3155 [gr-qc]} \BibitemShut
  {NoStop}%
%%CITATION = ARXIV:1402.3155;%%
\bibitem [{\citenamefont {Alesci}\ and\ \citenamefont
  {Cianfrani}(2015{\natexlab{a}})}]{QRLQGV}%
  \BibitemOpen
  \bibfield  {author} {\bibinfo {author} {\bibfnamefont {E.}~\bibnamefont
  {Alesci}}\ and\ \bibinfo {author} {\bibfnamefont {F.}~\bibnamefont
  {Cianfrani}},\ }\href {\doibase 10.1209/0295-5075/111/40002} {\bibfield
  {journal} {\bibinfo  {journal} {EPL}\ }\textbf {\bibinfo {volume} {111}},\
  \bibinfo {pages} {40002} (\bibinfo {year} {2015}{\natexlab{a}})},\ \Eprint
  {http://arxiv.org/abs/1410.4788} {arXiv:1410.4788 [gr-qc]} \BibitemShut
  {NoStop}%
%%CITATION = ARXIV:1410.4788;%%
\bibitem [{\citenamefont {Alesci}\ and\ \citenamefont
  {Cianfrani}(2015{\natexlab{b}})}]{QRLQGVI}%
  \BibitemOpen
  \bibfield  {author} {\bibinfo {author} {\bibfnamefont {E.}~\bibnamefont
  {Alesci}}\ and\ \bibinfo {author} {\bibfnamefont {F.}~\bibnamefont
  {Cianfrani}},\ }\href {\doibase 10.1103/PhysRevD.92.084065} {\bibfield
  {journal} {\bibinfo  {journal} {Phys. Rev.}\ }\textbf {\bibinfo {volume}
  {D92}},\ \bibinfo {pages} {084065} (\bibinfo {year} {2015}{\natexlab{b}})},\
  \Eprint {http://arxiv.org/abs/1506.07835} {arXiv:1506.07835 [gr-qc]}
  \BibitemShut {NoStop}%
%%CITATION = ARXIV:1506.07835;%%
\bibitem [{\citenamefont {Alesci}\ and\ \citenamefont
  {Cianfrani}(2016)}]{QRLQGVII}%
  \BibitemOpen
  \bibfield  {author} {\bibinfo {author} {\bibfnamefont {E.}~\bibnamefont
  {Alesci}}\ and\ \bibinfo {author} {\bibfnamefont {F.}~\bibnamefont
  {Cianfrani}},\ }\href {\doibase 10.1142/S0218271816420050} {\bibfield
  {journal} {\bibinfo  {journal} {Int. J. Mod. Phys.}\ }\textbf {\bibinfo
  {volume} {D25}},\ \bibinfo {pages} {1642005} (\bibinfo {year} {2016})},\
  \Eprint {http://arxiv.org/abs/1602.05475} {arXiv:1602.05475 [gr-qc]}
  \BibitemShut {NoStop}%
%%CITATION = ARXIV:1602.05475;%%
\bibitem [{\citenamefont {Oriti}\ \emph {et~al.}(2016)\citenamefont {Oriti},
  \citenamefont {Sindoni},\ and\ \citenamefont {Wilson-Ewing}}]{CosmoGFTI}%
  \BibitemOpen
  \bibfield  {author} {\bibinfo {author} {\bibfnamefont {D.}~\bibnamefont
  {Oriti}}, \bibinfo {author} {\bibfnamefont {L.}~\bibnamefont {Sindoni}}, \
  and\ \bibinfo {author} {\bibfnamefont {E.}~\bibnamefont {Wilson-Ewing}},\
  }\href {\doibase 10.1088/0264-9381/33/22/224001} {\bibfield  {journal}
  {\bibinfo  {journal} {Class. Quant. Grav.}\ }\textbf {\bibinfo {volume}
  {33}},\ \bibinfo {pages} {224001} (\bibinfo {year} {2016})},\ \Eprint
  {http://arxiv.org/abs/1602.05881} {arXiv:1602.05881 [gr-qc]} \BibitemShut
  {NoStop}%
%%CITATION = ARXIV:1602.05881;%%
\bibitem [{\citenamefont {Oriti}\ \emph {et~al.}(2017)\citenamefont {Oriti},
  \citenamefont {Sindoni},\ and\ \citenamefont {Wilson-Ewing}}]{CosmoGFTII}%
  \BibitemOpen
  \bibfield  {author} {\bibinfo {author} {\bibfnamefont {D.}~\bibnamefont
  {Oriti}}, \bibinfo {author} {\bibfnamefont {L.}~\bibnamefont {Sindoni}}, \
  and\ \bibinfo {author} {\bibfnamefont {E.}~\bibnamefont {Wilson-Ewing}},\
  }\href {\doibase 10.1088/1361-6382/aa549a} {\bibfield  {journal} {\bibinfo
  {journal} {Class. Quant. Grav.}\ }\textbf {\bibinfo {volume} {34}},\ \bibinfo
  {pages} {04LT01} (\bibinfo {year} {2017})},\ \Eprint
  {http://arxiv.org/abs/1602.08271} {arXiv:1602.08271 [gr-qc]} \BibitemShut
  {NoStop}%
%%CITATION = ARXIV:1602.08271;%%
\bibitem [{\citenamefont {Dapor}\ and\ \citenamefont
  {Liegener}(2018{\natexlab{a}})}]{DLI}%
  \BibitemOpen
  \bibfield  {author} {\bibinfo {author} {\bibfnamefont {A.}~\bibnamefont
  {Dapor}}\ and\ \bibinfo {author} {\bibfnamefont {K.}~\bibnamefont
  {Liegener}},\ }\href {\doibase 10.1016/j.physletb.2018.09.005} {\bibfield
  {journal} {\bibinfo  {journal} {Phys. Lett.}\ }\textbf {\bibinfo {volume}
  {B785}},\ \bibinfo {pages} {506} (\bibinfo {year} {2018}{\natexlab{a}})},\
  \Eprint {http://arxiv.org/abs/1706.09833} {arXiv:1706.09833 [gr-qc]}
  \BibitemShut {NoStop}%
%%CITATION = ARXIV:1706.09833;%%
\bibitem [{\citenamefont {Dapor}\ and\ \citenamefont
  {Liegener}(2018{\natexlab{b}})}]{DLII}%
  \BibitemOpen
  \bibfield  {author} {\bibinfo {author} {\bibfnamefont {A.}~\bibnamefont
  {Dapor}}\ and\ \bibinfo {author} {\bibfnamefont {K.}~\bibnamefont
  {Liegener}},\ }\href {\doibase 10.1088/1361-6382/aac4ba} {\bibfield
  {journal} {\bibinfo  {journal} {Class. Quant. Grav.}\ }\textbf {\bibinfo
  {volume} {35}},\ \bibinfo {pages} {135011} (\bibinfo {year}
  {2018}{\natexlab{b}})},\ \Eprint {http://arxiv.org/abs/1710.04015}
  {arXiv:1710.04015 [gr-qc]} \BibitemShut {NoStop}%
%%CITATION = ARXIV:1710.04015;%%
\bibitem [{\citenamefont {Han}\ and\ \citenamefont
  {Liu}(2020)}]{HanEffectiveDynamics}%
  \BibitemOpen
  \bibfield  {author} {\bibinfo {author} {\bibfnamefont {M.}~\bibnamefont
  {Han}}\ and\ \bibinfo {author} {\bibfnamefont {H.}~\bibnamefont {Liu}},\
  }\href {\doibase 10.1103/PhysRevD.101.046003} {\bibfield  {journal} {\bibinfo
   {journal} {Phys. Rev.}\ }\textbf {\bibinfo {volume} {D101}},\ \bibinfo
  {pages} {046003} (\bibinfo {year} {2020})},\ \Eprint
  {http://arxiv.org/abs/1910.03763} {arXiv:1910.03763 [gr-qc]} \BibitemShut
  {NoStop}%
%%CITATION = ARXIV:1910.03763;%%
\bibitem [{\citenamefont {Kisielowski}(2020)}]{HomogeneousIsotropic}%
  \BibitemOpen
  \bibfield  {author} {\bibinfo {author} {\bibfnamefont {M.}~\bibnamefont
  {Kisielowski}},\ }\href {\doibase 10.1088/1361-6382/ab9bb9} {\bibfield
  {journal} {\bibinfo  {journal} {Class. Quant. Grav.}\ }\textbf {\bibinfo
  {volume} {37}},\ \bibinfo {pages} {185004} (\bibinfo {year} {2020})},\
  \Eprint {http://arxiv.org/abs/1911.09639} {arXiv:1911.09639 [gr-qc]}
  \BibitemShut {NoStop}%
\bibitem [{\citenamefont {Lewandowski}\ and\ \citenamefont
  {Sahlmann}(2015)}]{LewandowskiSahlmann}%
  \BibitemOpen
  \bibfield  {author} {\bibinfo {author} {\bibfnamefont {J.}~\bibnamefont
  {Lewandowski}}\ and\ \bibinfo {author} {\bibfnamefont {H.}~\bibnamefont
  {Sahlmann}},\ }\href {\doibase 10.1103/PhysRevD.91.044022} {\bibfield
  {journal} {\bibinfo  {journal} {Phys. Rev.}\ }\textbf {\bibinfo {volume}
  {D91}},\ \bibinfo {pages} {044022} (\bibinfo {year} {2015})},\ \Eprint
  {http://arxiv.org/abs/1410.5276} {arXiv:1410.5276 [gr-qc]} \BibitemShut
  {NoStop}%
%%CITATION = ARXIV:1410.5276;%%
\bibitem [{\citenamefont {Alesci}\ \emph {et~al.}(2015)\citenamefont {Alesci},
  \citenamefont {Assanioussi}, \citenamefont {Lewandowski},\ and\ \citenamefont
  {Mäkinen}}]{HamiltonianOperator}%
  \BibitemOpen
  \bibfield  {author} {\bibinfo {author} {\bibfnamefont {E.}~\bibnamefont
  {Alesci}}, \bibinfo {author} {\bibfnamefont {M.}~\bibnamefont {Assanioussi}},
  \bibinfo {author} {\bibfnamefont {J.}~\bibnamefont {Lewandowski}}, \ and\
  \bibinfo {author} {\bibfnamefont {I.}~\bibnamefont {Mäkinen}},\ }\href
  {\doibase 10.1103/PhysRevD.91.124067} {\bibfield  {journal} {\bibinfo
  {journal} {Phys. Rev.}\ }\textbf {\bibinfo {volume} {D91}},\ \bibinfo {pages}
  {124067} (\bibinfo {year} {2015})},\ \Eprint
  {http://arxiv.org/abs/1504.02068} {arXiv:1504.02068 [gr-qc]} \BibitemShut
  {NoStop}%
%%CITATION = ARXIV:1504.02068;%%
\bibitem [{\citenamefont {Assanioussi}\ \emph {et~al.}(2015)\citenamefont
  {Assanioussi}, \citenamefont {Lewandowski},\ and\ \citenamefont
  {Mäkinen}}]{NewScalarConstraint}%
  \BibitemOpen
  \bibfield  {author} {\bibinfo {author} {\bibfnamefont {M.}~\bibnamefont
  {Assanioussi}}, \bibinfo {author} {\bibfnamefont {J.}~\bibnamefont
  {Lewandowski}}, \ and\ \bibinfo {author} {\bibfnamefont {I.}~\bibnamefont
  {Mäkinen}},\ }\href {\doibase 10.1103/PhysRevD.92.044042} {\bibfield
  {journal} {\bibinfo  {journal} {Phys. Rev.}\ }\textbf {\bibinfo {volume}
  {D92}},\ \bibinfo {pages} {044042} (\bibinfo {year} {2015})},\ \Eprint
  {http://arxiv.org/abs/1506.00299} {arXiv:1506.00299 [gr-qc]} \BibitemShut
  {NoStop}%
%%CITATION = ARXIV:1506.00299;%%
\bibitem [{\citenamefont {Assanioussi}\ \emph {et~al.}(2017)\citenamefont
  {Assanioussi}, \citenamefont {Lewandowski},\ and\ \citenamefont
  {Mäkinen}}]{TimeEvolution}%
  \BibitemOpen
  \bibfield  {author} {\bibinfo {author} {\bibfnamefont {M.}~\bibnamefont
  {Assanioussi}}, \bibinfo {author} {\bibfnamefont {J.}~\bibnamefont
  {Lewandowski}}, \ and\ \bibinfo {author} {\bibfnamefont {I.}~\bibnamefont
  {Mäkinen}},\ }\href {\doibase 10.1103/PhysRevD.96.024043} {\bibfield
  {journal} {\bibinfo  {journal} {Phys. Rev.}\ }\textbf {\bibinfo {volume}
  {D96}},\ \bibinfo {pages} {024043} (\bibinfo {year} {2017})},\ \Eprint
  {http://arxiv.org/abs/1702.01688} {arXiv:1702.01688 [gr-qc]} \BibitemShut
  {NoStop}%
%%CITATION = ARXIV:1702.01688;%%
\bibitem [{\citenamefont {Kisielowski}(2021)}]{VolumeMonochromatic}%
  \BibitemOpen
  \bibfield  {author} {\bibinfo {author} {\bibfnamefont {M.}~\bibnamefont
  {Kisielowski}},\ }\href {\doibase 10.1088/1361-6382/ac1e70} {\bibfield
  {journal} {\bibinfo  {journal} {Class. Quant. Grav.}\ }\textbf {\bibinfo
  {volume} {38}},\ \bibinfo {pages} {195023} (\bibinfo {year} {2021})},\
  \Eprint {http://arxiv.org/abs/2104.11010} {arXiv:2104.11010 [gr-qc]}
  \BibitemShut {NoStop}%
\bibitem [{\citenamefont {Rovelli}\ and\ \citenamefont
  {Smolin}(1995)}]{RovelliSmolinVolume}%
  \BibitemOpen
  \bibfield  {author} {\bibinfo {author} {\bibfnamefont {C.}~\bibnamefont
  {Rovelli}}\ and\ \bibinfo {author} {\bibfnamefont {L.}~\bibnamefont
  {Smolin}},\ }\href {\doibase 10.1016/0550-3213(95)00150-Q} {\bibfield
  {journal} {\bibinfo  {journal} {Nucl. Phys. B}\ }\textbf {\bibinfo {volume}
  {442}},\ \bibinfo {pages} {593} (\bibinfo {year} {1995})},\ \bibinfo {note}
  {[Erratum: Nucl.Phys.B 456, 753--754 (1995)]},\ \Eprint
  {http://arxiv.org/abs/gr-qc/9411005} {arXiv:gr-qc/9411005} \BibitemShut
  {NoStop}%
\bibitem [{\citenamefont {Perelomov}(1986)}]{Perelomov1986}%
  \BibitemOpen
  \bibfield  {author} {\bibinfo {author} {\bibfnamefont {A.}~\bibnamefont
  {Perelomov}},\ }\enquote {\bibinfo {title} {Coherent states for the rotation
  group of three-dimensional space},}\ in\ \href {\doibase
  10.1007/978-3-642-61629-7_5} {\emph {\bibinfo {booktitle} {Generalized
  Coherent States and Their Applications}}}\ (\bibinfo  {publisher} {Springer
  Berlin Heidelberg},\ \bibinfo {address} {Berlin, Heidelberg},\ \bibinfo
  {year} {1986})\ pp.\ \bibinfo {pages} {54--66}\BibitemShut {NoStop}%
\bibitem [{\citenamefont {Major}(1999)}]{MajorQuantizedDirection}%
  \BibitemOpen
  \bibfield  {author} {\bibinfo {author} {\bibfnamefont {S.~A.}\ \bibnamefont
  {Major}},\ }\href {\doibase 10.1088/0264-9381/16/12/307} {\bibfield
  {journal} {\bibinfo  {journal} {Class. Quant. Grav.}\ }\textbf {\bibinfo
  {volume} {16}},\ \bibinfo {pages} {3859} (\bibinfo {year} {1999})},\ \Eprint
  {http://arxiv.org/abs/gr-qc/9905019} {arXiv:gr-qc/9905019} \BibitemShut
  {NoStop}%
\bibitem [{\citenamefont {Major}\ and\ \citenamefont
  {Seifert}(2002)}]{MajorSeifert}%
  \BibitemOpen
  \bibfield  {author} {\bibinfo {author} {\bibfnamefont {S.~A.}\ \bibnamefont
  {Major}}\ and\ \bibinfo {author} {\bibfnamefont {M.~D.}\ \bibnamefont
  {Seifert}},\ }\href {\doibase 10.1088/0264-9381/19/8/311} {\bibfield
  {journal} {\bibinfo  {journal} {Class. Quant. Grav.}\ }\textbf {\bibinfo
  {volume} {19}},\ \bibinfo {pages} {2211} (\bibinfo {year} {2002})},\ \Eprint
  {http://arxiv.org/abs/gr-qc/0109056} {arXiv:gr-qc/0109056} \BibitemShut
  {NoStop}%
\bibitem [{\citenamefont {Bianchi}\ \emph {et~al.}(2010)\citenamefont
  {Bianchi}, \citenamefont {Rovelli},\ and\ \citenamefont {Vidotto}}]{BRV}%
  \BibitemOpen
  \bibfield  {author} {\bibinfo {author} {\bibfnamefont {E.}~\bibnamefont
  {Bianchi}}, \bibinfo {author} {\bibfnamefont {C.}~\bibnamefont {Rovelli}}, \
  and\ \bibinfo {author} {\bibfnamefont {F.}~\bibnamefont {Vidotto}},\ }\href
  {\doibase 10.1103/PhysRevD.82.084035} {\bibfield  {journal} {\bibinfo
  {journal} {Phys.Rev.}\ }\textbf {\bibinfo {volume} {D82}},\ \bibinfo {pages}
  {084035} (\bibinfo {year} {2010})},\ \Eprint {http://arxiv.org/abs/1003.3483}
  {arXiv:1003.3483 [gr-qc]} \BibitemShut {NoStop}%
%%CITATION = ARXIV:1003.3483;%%
\bibitem [{\citenamefont {Barrett}\ \emph {et~al.}(2009)\citenamefont
  {Barrett}, \citenamefont {Dowdall}, \citenamefont {Fairbairn}, \citenamefont
  {Gomes},\ and\ \citenamefont {Hellmann}}]{EuclideanEPRLAsymptotics}%
  \BibitemOpen
  \bibfield  {author} {\bibinfo {author} {\bibfnamefont {J.~W.}\ \bibnamefont
  {Barrett}}, \bibinfo {author} {\bibfnamefont {R.}~\bibnamefont {Dowdall}},
  \bibinfo {author} {\bibfnamefont {W.~J.}\ \bibnamefont {Fairbairn}}, \bibinfo
  {author} {\bibfnamefont {H.}~\bibnamefont {Gomes}}, \ and\ \bibinfo {author}
  {\bibfnamefont {F.}~\bibnamefont {Hellmann}},\ }\href {\doibase
  10.1063/1.3244218} {\bibfield  {journal} {\bibinfo  {journal} {J.Math.Phys.}\
  }\textbf {\bibinfo {volume} {50}},\ \bibinfo {pages} {112504} (\bibinfo
  {year} {2009})},\ \Eprint {http://arxiv.org/abs/0902.1170} {arXiv:0902.1170
  [gr-qc]} \BibitemShut {NoStop}%
%%CITATION = ARXIV:0902.1170;%%
\bibitem [{\citenamefont {Livine}\ and\ \citenamefont {Speziale}(2007)}]{LSiI}%
  \BibitemOpen
  \bibfield  {author} {\bibinfo {author} {\bibfnamefont {E.~R.}\ \bibnamefont
  {Livine}}\ and\ \bibinfo {author} {\bibfnamefont {S.}~\bibnamefont
  {Speziale}},\ }\href {\doibase 10.1103/PhysRevD.76.084028} {\bibfield
  {journal} {\bibinfo  {journal} {Phys.Rev.}\ }\textbf {\bibinfo {volume}
  {D76}},\ \bibinfo {pages} {084028} (\bibinfo {year} {2007})},\ \Eprint
  {http://arxiv.org/abs/0705.0674} {arXiv:0705.0674 [gr-qc]} \BibitemShut
  {NoStop}%
%%CITATION = ARXIV:0705.0674;%%
\bibitem [{\citenamefont {Kisielowski}\ and\ \citenamefont
  {Lewandowski}(2019)}]{SFscalar}%
  \BibitemOpen
  \bibfield  {author} {\bibinfo {author} {\bibfnamefont {M.}~\bibnamefont
  {Kisielowski}}\ and\ \bibinfo {author} {\bibfnamefont {J.}~\bibnamefont
  {Lewandowski}},\ }\href {\doibase 10.1088/1361-6382/aafcc0} {\bibfield
  {journal} {\bibinfo  {journal} {Class. Quant. Grav.}\ }\textbf {\bibinfo
  {volume} {36}},\ \bibinfo {pages} {075006} (\bibinfo {year} {2019})},\
  \Eprint {http://arxiv.org/abs/1807.06098} {arXiv:1807.06098 [gr-qc]}
  \BibitemShut {NoStop}%
%%CITATION = ARXIV:1807.06098;%%
\bibitem [{\citenamefont {Zhang}\ \emph {et~al.}(2019)\citenamefont {Zhang},
  \citenamefont {Lewandowski}, \citenamefont {Li},\ and\ \citenamefont
  {Ma}}]{ZhangBouncingUniverse}%
  \BibitemOpen
  \bibfield  {author} {\bibinfo {author} {\bibfnamefont {C.}~\bibnamefont
  {Zhang}}, \bibinfo {author} {\bibfnamefont {J.}~\bibnamefont {Lewandowski}},
  \bibinfo {author} {\bibfnamefont {H.}~\bibnamefont {Li}}, \ and\ \bibinfo
  {author} {\bibfnamefont {Y.}~\bibnamefont {Ma}},\ }\href {\doibase
  10.1103/PhysRevD.99.124012} {\bibfield  {journal} {\bibinfo  {journal} {Phys.
  Rev. D}\ }\textbf {\bibinfo {volume} {99}},\ \bibinfo {pages} {124012}
  (\bibinfo {year} {2019})},\ \Eprint {http://arxiv.org/abs/1904.07046}
  {arXiv:1904.07046 [gr-qc]} \BibitemShut {NoStop}%
\bibitem [{\citenamefont {Ashtekar}\ \emph {et~al.}(2007)\citenamefont
  {Ashtekar}, \citenamefont {Pawlowski}, \citenamefont {Singh},\ and\
  \citenamefont {Vandersloot}}]{AshtekarPawlowskiClosedUniverse}%
  \BibitemOpen
  \bibfield  {author} {\bibinfo {author} {\bibfnamefont {A.}~\bibnamefont
  {Ashtekar}}, \bibinfo {author} {\bibfnamefont {T.}~\bibnamefont {Pawlowski}},
  \bibinfo {author} {\bibfnamefont {P.}~\bibnamefont {Singh}}, \ and\ \bibinfo
  {author} {\bibfnamefont {K.}~\bibnamefont {Vandersloot}},\ }\href {\doibase
  10.1103/PhysRevD.75.024035} {\bibfield  {journal} {\bibinfo  {journal} {Phys.
  Rev. D}\ }\textbf {\bibinfo {volume} {75}},\ \bibinfo {pages} {024035}
  (\bibinfo {year} {2007})},\ \Eprint {http://arxiv.org/abs/gr-qc/0612104}
  {arXiv:gr-qc/0612104} \BibitemShut {NoStop}%
\bibitem [{\citenamefont {Pawlowski}\ and\ \citenamefont
  {Ashtekar}(2012)}]{AshtekarPawlowskiCosmologicalConstant}%
  \BibitemOpen
  \bibfield  {author} {\bibinfo {author} {\bibfnamefont {T.}~\bibnamefont
  {Pawlowski}}\ and\ \bibinfo {author} {\bibfnamefont {A.}~\bibnamefont
  {Ashtekar}},\ }\href {\doibase 10.1103/PhysRevD.85.064001} {\bibfield
  {journal} {\bibinfo  {journal} {Phys. Rev. D}\ }\textbf {\bibinfo {volume}
  {85}},\ \bibinfo {pages} {064001} (\bibinfo {year} {2012})},\ \Eprint
  {http://arxiv.org/abs/1112.0360} {arXiv:1112.0360 [gr-qc]} \BibitemShut
  {NoStop}%
\bibitem [{\citenamefont {Beetle}\ \emph {et~al.}(2016)\citenamefont {Beetle},
  \citenamefont {Engle}, \citenamefont {Hogan},\ and\ \citenamefont
  {Mendonca}}]{EngleLQCI}%
  \BibitemOpen
  \bibfield  {author} {\bibinfo {author} {\bibfnamefont {C.}~\bibnamefont
  {Beetle}}, \bibinfo {author} {\bibfnamefont {J.~S.}\ \bibnamefont {Engle}},
  \bibinfo {author} {\bibfnamefont {M.~E.}\ \bibnamefont {Hogan}}, \ and\
  \bibinfo {author} {\bibfnamefont {P.}~\bibnamefont {Mendonca}},\ }\href
  {\doibase 10.1142/S0218271816420128} {\bibfield  {journal} {\bibinfo
  {journal} {Int. J. Mod. Phys. D}\ }\textbf {\bibinfo {volume} {25}},\
  \bibinfo {pages} {1642012} (\bibinfo {year} {2016})},\ \Eprint
  {http://arxiv.org/abs/1603.01128} {arXiv:1603.01128 [gr-qc]} \BibitemShut
  {NoStop}%
\bibitem [{\citenamefont {Beetle}\ \emph {et~al.}(2017)\citenamefont {Beetle},
  \citenamefont {Engle}, \citenamefont {Hogan},\ and\ \citenamefont
  {Mendon\c{c}a}}]{EngleLQCII}%
  \BibitemOpen
  \bibfield  {author} {\bibinfo {author} {\bibfnamefont {C.}~\bibnamefont
  {Beetle}}, \bibinfo {author} {\bibfnamefont {J.~S.}\ \bibnamefont {Engle}},
  \bibinfo {author} {\bibfnamefont {M.~E.}\ \bibnamefont {Hogan}}, \ and\
  \bibinfo {author} {\bibfnamefont {P.}~\bibnamefont {Mendon\c{c}a}},\ }\href
  {\doibase 10.1088/1361-6382/aa89c6} {\bibfield  {journal} {\bibinfo
  {journal} {Class. Quant. Grav.}\ }\textbf {\bibinfo {volume} {34}},\ \bibinfo
  {pages} {225009} (\bibinfo {year} {2017})},\ \Eprint
  {http://arxiv.org/abs/1706.02424} {arXiv:1706.02424 [gr-qc]} \BibitemShut
  {NoStop}%
\bibitem [{\citenamefont {Engle}\ and\ \citenamefont
  {Vilensky}(2018)}]{EngleLQCIII}%
  \BibitemOpen
  \bibfield  {author} {\bibinfo {author} {\bibfnamefont {J.}~\bibnamefont
  {Engle}}\ and\ \bibinfo {author} {\bibfnamefont {I.}~\bibnamefont
  {Vilensky}},\ }\href {\doibase 10.1103/PhysRevD.98.023505} {\bibfield
  {journal} {\bibinfo  {journal} {Phys. Rev. D}\ }\textbf {\bibinfo {volume}
  {98}},\ \bibinfo {pages} {023505} (\bibinfo {year} {2018})},\ \Eprint
  {http://arxiv.org/abs/1802.01543} {arXiv:1802.01543 [gr-qc]} \BibitemShut
  {NoStop}%
\bibitem [{\citenamefont {Ashtekar}\ and\ \citenamefont
  {Gupt}(2017{\natexlab{a}})}]{AshtekarCMBI}%
  \BibitemOpen
  \bibfield  {author} {\bibinfo {author} {\bibfnamefont {A.}~\bibnamefont
  {Ashtekar}}\ and\ \bibinfo {author} {\bibfnamefont {B.}~\bibnamefont
  {Gupt}},\ }\href {\doibase 10.1088/1361-6382/34/1/014002} {\bibfield
  {journal} {\bibinfo  {journal} {Class. Quant. Grav.}\ }\textbf {\bibinfo
  {volume} {34}},\ \bibinfo {pages} {014002} (\bibinfo {year}
  {2017}{\natexlab{a}})},\ \Eprint {http://arxiv.org/abs/1608.04228}
  {arXiv:1608.04228 [gr-qc]} \BibitemShut {NoStop}%
\bibitem [{\citenamefont {Ashtekar}\ and\ \citenamefont
  {Gupt}(2017{\natexlab{b}})}]{AshtekarCMBII}%
  \BibitemOpen
  \bibfield  {author} {\bibinfo {author} {\bibfnamefont {A.}~\bibnamefont
  {Ashtekar}}\ and\ \bibinfo {author} {\bibfnamefont {B.}~\bibnamefont
  {Gupt}},\ }\href {\doibase 10.1088/1361-6382/aa52d4} {\bibfield  {journal}
  {\bibinfo  {journal} {Class. Quant. Grav.}\ }\textbf {\bibinfo {volume}
  {34}},\ \bibinfo {pages} {035004} (\bibinfo {year} {2017}{\natexlab{b}})},\
  \Eprint {http://arxiv.org/abs/1610.09424} {arXiv:1610.09424 [gr-qc]}
  \BibitemShut {NoStop}%
\bibitem [{\citenamefont {Agullo}\ \emph {et~al.}(2017)\citenamefont {Agullo},
  \citenamefont {Ashtekar},\ and\ \citenamefont {Gupt}}]{AshtekarCMBIII}%
  \BibitemOpen
  \bibfield  {author} {\bibinfo {author} {\bibfnamefont {I.}~\bibnamefont
  {Agullo}}, \bibinfo {author} {\bibfnamefont {A.}~\bibnamefont {Ashtekar}}, \
  and\ \bibinfo {author} {\bibfnamefont {B.}~\bibnamefont {Gupt}},\ }\href
  {\doibase 10.1088/1361-6382/aa60ec} {\bibfield  {journal} {\bibinfo
  {journal} {Class. Quant. Grav.}\ }\textbf {\bibinfo {volume} {34}},\ \bibinfo
  {pages} {074003} (\bibinfo {year} {2017})},\ \Eprint
  {http://arxiv.org/abs/1611.09810} {arXiv:1611.09810 [gr-qc]} \BibitemShut
  {NoStop}%
\bibitem [{\citenamefont {Ashtekar}\ \emph {et~al.}(2009)\citenamefont
  {Ashtekar}, \citenamefont {Campiglia},\ and\ \citenamefont
  {Henderson}}]{AshtekarSpinFoamI}%
  \BibitemOpen
  \bibfield  {author} {\bibinfo {author} {\bibfnamefont {A.}~\bibnamefont
  {Ashtekar}}, \bibinfo {author} {\bibfnamefont {M.}~\bibnamefont {Campiglia}},
  \ and\ \bibinfo {author} {\bibfnamefont {A.}~\bibnamefont {Henderson}},\
  }\href {\doibase 10.1016/j.physletb.2009.10.042} {\bibfield  {journal}
  {\bibinfo  {journal} {Phys.Lett.}\ }\textbf {\bibinfo {volume} {B681}},\
  \bibinfo {pages} {347} (\bibinfo {year} {2009})},\ \Eprint
  {http://arxiv.org/abs/0909.4221} {arXiv:0909.4221 [gr-qc]} \BibitemShut
  {NoStop}%
%%CITATION = ARXIV:0909.4221;%%
\bibitem [{\citenamefont {Ashtekar}\ \emph {et~al.}(2010)\citenamefont
  {Ashtekar}, \citenamefont {Campiglia},\ and\ \citenamefont
  {Henderson}}]{AshtekarSpinFoamII}%
  \BibitemOpen
  \bibfield  {author} {\bibinfo {author} {\bibfnamefont {A.}~\bibnamefont
  {Ashtekar}}, \bibinfo {author} {\bibfnamefont {M.}~\bibnamefont {Campiglia}},
  \ and\ \bibinfo {author} {\bibfnamefont {A.}~\bibnamefont {Henderson}},\
  }\href {\doibase 10.1088/0264-9381/27/13/135020} {\bibfield  {journal}
  {\bibinfo  {journal} {Class.Quant.Grav.}\ }\textbf {\bibinfo {volume} {27}},\
  \bibinfo {pages} {135020} (\bibinfo {year} {2010})},\ \Eprint
  {http://arxiv.org/abs/1001.5147} {arXiv:1001.5147 [gr-qc]} \BibitemShut
  {NoStop}%
%%CITATION = ARXIV:1001.5147;%%
\bibitem [{\citenamefont {Campiglia}\ \emph {et~al.}(2010)\citenamefont
  {Campiglia}, \citenamefont {Henderson},\ and\ \citenamefont
  {Nelson}}]{AshtekarSpinFoamIII}%
  \BibitemOpen
  \bibfield  {author} {\bibinfo {author} {\bibfnamefont {M.}~\bibnamefont
  {Campiglia}}, \bibinfo {author} {\bibfnamefont {A.}~\bibnamefont
  {Henderson}}, \ and\ \bibinfo {author} {\bibfnamefont {W.}~\bibnamefont
  {Nelson}},\ }\href {\doibase 10.1103/PhysRevD.82.064036} {\bibfield
  {journal} {\bibinfo  {journal} {Phys.Rev.}\ }\textbf {\bibinfo {volume}
  {D82}},\ \bibinfo {pages} {064036} (\bibinfo {year} {2010})},\ \Eprint
  {http://arxiv.org/abs/1007.3723} {arXiv:1007.3723 [gr-qc]} \BibitemShut
  {NoStop}%
%%CITATION = ARXIV:1007.3723;%%
\bibitem [{\citenamefont {Ashtekar}\ and\ \citenamefont
  {Lewandowski}(1998)}]{AshtekarLewandowskiVolume}%
  \BibitemOpen
  \bibfield  {author} {\bibinfo {author} {\bibfnamefont {A.}~\bibnamefont
  {Ashtekar}}\ and\ \bibinfo {author} {\bibfnamefont {J.}~\bibnamefont
  {Lewandowski}},\ }\href {\doibase 10.4310/ATMP.1997.v1.n2.a8} {\bibfield
  {journal} {\bibinfo  {journal} {Adv. Theor. Math. Phys.}\ }\textbf {\bibinfo
  {volume} {1}},\ \bibinfo {pages} {388} (\bibinfo {year} {1998})},\ \Eprint
  {http://arxiv.org/abs/gr-qc/9711031} {arXiv:gr-qc/9711031} \BibitemShut
  {NoStop}%
\bibitem [{\citenamefont {Giesel}\ and\ \citenamefont
  {Thiemann}(2006{\natexlab{a}})}]{ThiemannConsisencyCheckI}%
  \BibitemOpen
  \bibfield  {author} {\bibinfo {author} {\bibfnamefont {K.}~\bibnamefont
  {Giesel}}\ and\ \bibinfo {author} {\bibfnamefont {T.}~\bibnamefont
  {Thiemann}},\ }\href {\doibase 10.1088/0264-9381/23/18/011} {\bibfield
  {journal} {\bibinfo  {journal} {Class. Quant. Grav.}\ }\textbf {\bibinfo
  {volume} {23}},\ \bibinfo {pages} {5667} (\bibinfo {year}
  {2006}{\natexlab{a}})},\ \Eprint {http://arxiv.org/abs/gr-qc/0507036}
  {arXiv:gr-qc/0507036} \BibitemShut {NoStop}%
\bibitem [{\citenamefont {Giesel}\ and\ \citenamefont
  {Thiemann}(2006{\natexlab{b}})}]{ThiemannConsisencyCheckII}%
  \BibitemOpen
  \bibfield  {author} {\bibinfo {author} {\bibfnamefont {K.}~\bibnamefont
  {Giesel}}\ and\ \bibinfo {author} {\bibfnamefont {T.}~\bibnamefont
  {Thiemann}},\ }\href {\doibase 10.1088/0264-9381/23/18/012} {\bibfield
  {journal} {\bibinfo  {journal} {Class. Quant. Grav.}\ }\textbf {\bibinfo
  {volume} {23}},\ \bibinfo {pages} {5693} (\bibinfo {year}
  {2006}{\natexlab{b}})},\ \Eprint {http://arxiv.org/abs/gr-qc/0507037}
  {arXiv:gr-qc/0507037} \BibitemShut {NoStop}%
\end{thebibliography}%
\end{document}